\documentclass[sigconf]{acmart}

\usepackage{bm}
\usepackage{url}
\usepackage{xcolor}
\usepackage{bbm}
\usepackage{graphicx}
\usepackage{makecell}
\usepackage[noend]{algorithmic}
\usepackage{caption}
\usepackage{amsmath}
\usepackage{diagbox}
\usepackage{multirow}
\usepackage{enumitem}
\usepackage{subcaption}
\usepackage{tabularx}
\usepackage{color}
\usepackage{longtable}
\usepackage{supertabular}

\usepackage{mdframed}
\newmdenv[linecolor=black, backgroundcolor=gray!10]{boxedalgorithm}
\usepackage[ruled,vlined]{algorithm2e}
\usepackage{fancyvrb}
\usepackage{xcolor}
\usepackage{xspace}
\usepackage{color} 
\usepackage{layouts}
\usepackage[utf8]{inputenc}
\usepackage{eqparbox}
\usepackage{etoolbox} 
\usepackage{pgfplots}
\usepackage{marginnote}
\usepackage{caption}
\usepackage{listings}
\newcommand\vldbdoi{XX.XX/XXX.XX}
\newcommand\vldbpages{XXX-XXX}
\newcommand\vldbvolume{14}
\newcommand\vldbissue{1}
\newcommand\vldbyear{2024}
\newcommand\vldbauthors{\authors}
\newcommand\vldbtitle{\shorttitle} 
\newcommand\vldbavailabilityurl{https://github.com/pzwupenn/CORDON/}
\newcommand\vldbpagestyle{plain} 

\newcounter{nalg}[section] 
\renewcommand{\thenalg}{\arabic{nalg}} 
\DeclareCaptionLabelFormat{algocaption}{Algorithm \thenalg} 

\lstnewenvironment{algo}[1][] 
{   
    \refstepcounter{nalg} 
    \captionsetup{labelformat=algocaption,labelsep=colon} 
    \lstset{ 
        mathescape=true,
        frame=tB,
        numbers=left, 
        numberstyle=\tiny,
        basicstyle=\scriptsize, 
        keywordstyle=\color{black}\bfseries\em,
        keywords={,input, output, return, datatype, function, in, if, else, foreach, while, begin, end, } 
        numbers=left,
        xleftmargin=.04\textwidth,
        #1 
    }
}
{}

\author{Peizhi Wu}
\affiliation{%
 \institution{University of Pennsylvania}
 \postcode{19104}
}
\email{pagewu@cis.upenn.edu}
\author{Ryan Marcus}
\affiliation{%
 \institution{University of Pennsylvania}
 \postcode{19104}
}
\email{rcmarcus@cis.upenn.edu}
\author{Zachary G. Ives}
\affiliation{%
 \institution{University of Pennsylvania}
 \postcode{19104}
}
\email{zives@cis.upenn.edu}

\definecolor{mycolor}{RGB}{255,0,0}

\newcommand{\eat}[1]{}

\newcommand{\name}{\mbox{$\mathsf{CORDON}$}\xspace}

\begin{document}

\title{Adding Domain Knowledge to Query-Driven Learned Databases}


\begin{abstract}
 In recent years, \emph{learned cardinality estimation} has emerged as an alternative to traditional query optimization methods: by training machine learning models over observed query performance, learned cardinality estimation techniques can accurately predict query cardinalities and costs --- accounting for skew, correlated predicates, and many other factors that traditional methods struggle to capture. However, query-driven learned cardinality estimators are dependent on sample workloads, requiring vast amounts of labeled queries. Further, we show that state-of-the-art query-driven techniques can make significant and unpredictable errors on queries that are outside the distribution of their training set. We show that these out-of-distribution errors can be mitigated by incorporating the \emph{domain knowledge} used in traditional query optimizers: \emph{constraints} on values and cardinalities (e.g., based on key-foreign-key relationships, range predicates, and more generally on inclusion and functional dependencies). We develop methods for \emph{semi-supervised} query-driven learned query optimization, based on constraints, and we experimentally demonstrate that such techniques can increase a learned query optimizer's accuracy in cardinality estimation, reduce the reliance on massive labeled queries, and improve the robustness of query end-to-end performance.
\end{abstract}

\maketitle

\eat{

\pagestyle{\vldbpagestyle}
\begingroup\small\noindent\raggedright\textbf{PVLDB Reference Format:}\\
\vldbauthors. \vldbtitle. PVLDB, \vldbvolume(\vldbissue): \vldbpages, \vldbyear.\\
\href{https://doi.org/\vldbdoi}{doi:\vldbdoi}
\endgroup
\begingroup
\renewcommand\thefootnote{}\footnote{\noindent
This work is licensed under the Creative Commons BY-NC-ND 4.0 International License. Visit \url{https://creativecommons.org/licenses/by-nc-nd/4.0/} to view a copy of this license. For any use beyond those covered by this license, obtain permission by emailing \href{mailto:info@vldb.org}{info@vldb.org}. Copyright is held by the owner/author(s). Publication rights licensed to the VLDB Endowment. \\
\raggedright Proceedings of the VLDB Endowment, Vol. \vldbvolume, No. \vldbissue\ %
ISSN 2150-8097. \\
\href{https://doi.org/\vldbdoi}{doi:\vldbdoi} \\
}\addtocounter{footnote}{-1}\endgroup

\ifdefempty{\vldbavailabilityurl}{}{
\vspace{.3cm}
\begingroup\small\noindent\raggedright\textbf{PVLDB Artifact Availability:}\\
The source code, data, and/or other artifacts have been made available at \url{\vldbavailabilityurl}.
\endgroup
}

}

\section{Introduction}

\begin{figure}[tb]
\centering
\begin{subfigure}{.235\textwidth}
\captionsetup{justification=centering}
\includegraphics[width=\textwidth]{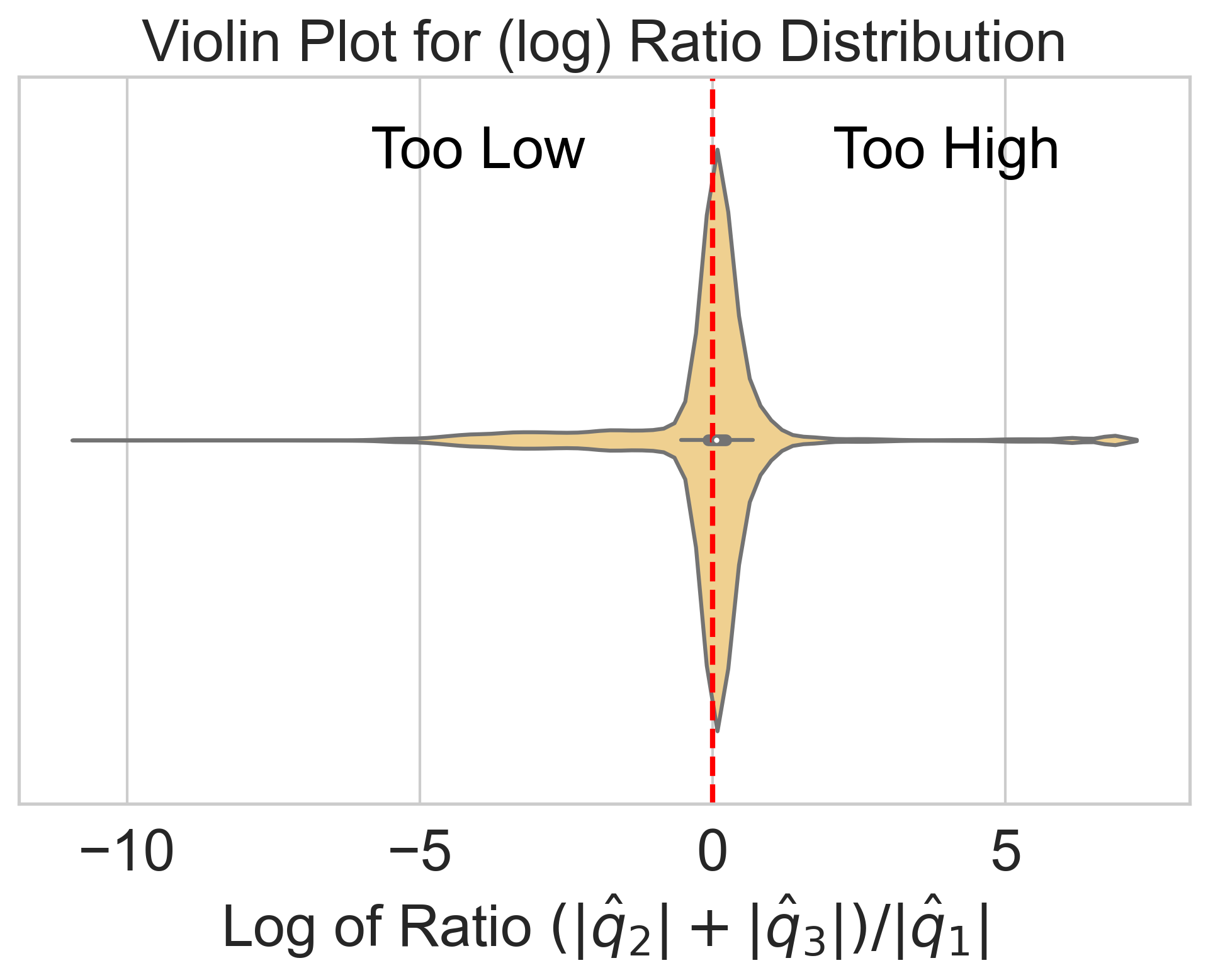}
\caption{Consistency Constraint:\\ ${|\hat q_2|}+{|\hat q_3|} = {|\hat q_1|}$ if $q_2 = q_1 - q_3$} \label{fig.example.cc}
\end{subfigure}
\begin{subfigure}{.235\textwidth}
\captionsetup{justification=centering}
\includegraphics[width=\textwidth]{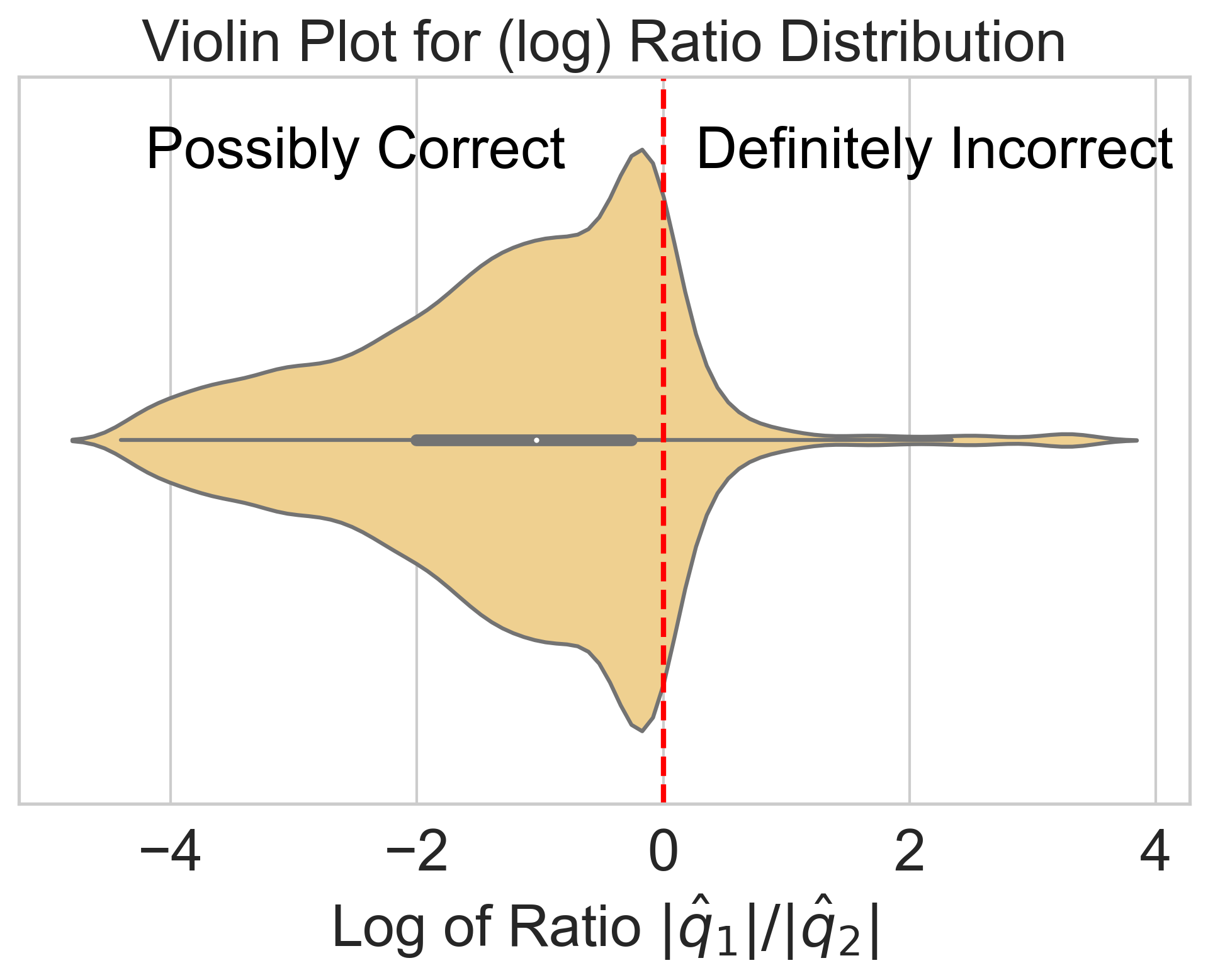}
\caption{PK-FK Inequality Constraint:\\ ${|\hat q_1|}\leq {|\hat q_2|}$ if $q_1 = q_2 \bowtie T$}       \label{fig.example.fk}
\end{subfigure}
\caption{Violin plots for MSCN predictions of two ratios of cardinality estimates ($\hat q_i$ for each query $q_i$) that should comply with database constraints. 
Left (a): consistency constraint over a range selection query that has been partitioned. Correct estimates are indicated by the red dashed line. The left and right sides of the line are too low and too high, respectively. Right (b): PK-FK inequality constraint over key-foreign key join queries. The left side of the red dashed line is possibly correct while the right side is definitely incorrect.
}\label{fig.example}
\end{figure}

Cardinality estimation has been a core problem~\cite{leis2015good} in data management research for decades~\cite{selinger1979access}.  A wide variety of cardinality estimation techniques have been developed, such as histograms, sketches, multidimensional histograms~\cite{bruno2001stholes}, sampling and synopses~\cite{garofalakis2001approximate}, statistics on query expressions~\cite{bruno2002exploiting}, learned compensation factors for query predicate estimates~\cite{markl2003leo}, and constraint-based bounds (e.g., key-foreign key dependencies~\cite{selinger1979access}.  More recently, learned cardinality estimators~\cite{kipf2019learned,local_card_est,hasan2020deep,negi2021flow,yang2020neurocard,deepdb}, have shown promise in replacing these often-complex or heuristic techniques with a single general framework using machine learning models.

Broadly, learned cardinality estimation techniques can be divided into \emph{data-driven} methods, such as DeepDB~\cite{deepdb}~and~NeuroCard~\cite{yang2020neurocard}, which use samples from a database to build a model of data distributions and seek to predict query cardinalities; and \emph{query-driven} methods such as Deep Sketches~\cite{deep_sketch}, which seek to take information from an executed sample workload and learn a  model to make accurate cardinality estimations. While data-driven cardinality estimators tend to produce accurate results for a wide variety of queries, query-driven cardinality estimators have a number of advantages, such as low model complexity and fast inference~\cite{negi2023robust}. One of the biggest downsides of query-driven cardinality estimators, however, is their reliance on a training set, which can (1) be expensive to collect, and (2) can result in query-driven models making poor predictions for queries that are highly dissimilar to those in the training set (i.e., ``out of distribution'' queries)~\cite{ready_for_lce}.

\paragraph{Query-driven learned cardinality estimators violate ``common sense'' constraints.} Surprisingly, query-driven learned cardinality estimators often make predictions that are not logically consistent with the underlying database. We illustrate two cases of this unexpected behavior in Figure~\ref{fig.example}: we train an MSCN~\cite{kipf2019learned} model to predict queries from DSB benchmark~\cite{ding2021dsb}, and test various queries with zero or one joins. First, in Figure~\ref{fig.example.cc}, we test to see if the predictions made by the MSCN model obey a simple consistency constraint. We do this by generating a range query, $q_1$, which selects all values between randomly generated points $a$ and $b$. We then pick a random midpoint $m$ such that $a < m < b$, and create (i) $q_2$, which scans all data from $a$ to $m$, and (ii) $q_3$, which scans all data from $m$ to $b$. Clearly, the cardinality of $q_1$ is equal to the cardinality of $q_2$ plus the cardinality of $q_3$, but the MSCN model could predict a cardinality for $q_2$ or $q_3$ that is orders of magnitude higher than the MSCN model's prediction for $q_1$!

Second, in Figure~\ref{fig.example.fk}, we test to see if the predictions made by the MSCN model obey a simple join cardinality constraint: the size of a PK-FK (primary key to foreign key) join of a fact table to a dimension table cannot exceed the size of the fact table. This knowledge was built into the very first cost-based query optimizers~\cite{selinger1979access}, yet surprisingly this simple observation tends to be poorly captured by learned cardinality estimation models: the MSCN model makes many predictions that clearly violate this PK-FK inequality constraint, and thus are definitely incorrect.

\paragraph{The need for domain knowledge} The failure of query-driven cardinality estimation techniques to learn such constraints should be especially troubling to data management researchers. Decades of database research have captured useful domain knowledge (i.e., semantic rules and intensional information) that aids in predictions. This knowledge can be looked at as conditions or facts that must be true in a database. Lines of database research have shown the value of such domain knowledge in many aspects of implementing and optimizing database systems, such as data integration~\cite{doan2012principles} and cardinality estimation~\cite{selinger1979access}. Despite the effectiveness of domain knowledge in traditional query processing, recent proposals for learned cardinality estimation models seldom (explicitly) take advantage of domain knowledge, which could be exceptionally useful for estimating cardinalities of ``out of distribution'' queries.


\smallskip
\textbf{Goal.}  The question we seek to answer in this paper is: can we design a general-purpose framework to effectively introduce database domain knowledge, like PK-FK constraints and query containment, into query-driven learned cardinality estimation systems?

\smallskip
\textbf{Solution.} Our key ideas in building such a framework for query-driven models is to \emph{use domain knowledge to generate differentiable constraints} that are represented by the label (output) relationships among different queries. We use these constraints to (1) \emph{modify the model's loss function to penalize constraint violations}, and (2) to \emph{augment the model's training set by generating new training instances from constraints}.  Importantly, the rest of the model's training procedure is entirely untouched: changes are only made in the form of additional loss function terms and additional training instances.
Together, these ideas comprise a \emph{semi-supervised approach} towards building more accurate query-driven cardinality estimation models. We call this framework \name (\textbf{C}onstraining \textbf{O}utput \textbf{R}anges with \textbf{D}omain \textbf{O}riented k\textbf{N}owledge).

\smallskip
\textbf{Contributions and key results.} To the best of our knowledge, \name proposes the first holistic learning framework for adding domain knowledge to query-driven learned cardinality estimation models. 

\begin{itemize}
\item{We identify and demonstrate that existing query-driven models do not consider domain knowledge, which could result in bad accuracy and poor generalization ability.}
\item{\name improves training sample efficiency, as \name can provide more training signals (\emph{loss function of constraints} and \emph{augmented training queries}) -- in some cases, \name requires \emph{half as much training data} to achieve the same accuracy as supervised techniques.}
\item{\name's domain knowledge creates better estimation results, \emph{improving cardinality estimation accuracy by up to two orders of magnitude on out-of-distribution queries}. Experimentally, we show that this improved estimation accuracy can improve query runtimes by 20-50\%.}
\item{\name is a general and extensible framework that allows users to add any domain knowledge in the form of the constraints defined by the paper.}
\end{itemize}


We hope the high-level ideas and the methodology of \name can inspire researchers and practitioners to revisit conventional and well-studied data management techniques when designing learned systems components. We believe that such considerations will result in significantly more robust designs.
\section{Preliminaries} \label{section.preliminaries}

\subsection{Query-Driven Learned Database Systems}
Traditional query optimization can be effective when it is based on summary structures on table attributes that have been computed offline (e.g., histograms, sketches). Unfortunately, correlated predicates, especially over skewed data, are difficult to predict with single-attribute synopses, often resulting in inaccurate estimates and poor plans~\cite{ioannidis1991propagation}.  This motivated research into multi-attribute structures such as join synopses~\cite{DBLP:conf/sigmod/AcharyaGPR99a} and multidimensional histograms~\cite{aboulnaga1999self}.
Yet these structures imposed significant up-front costs, motivating later study of \emph{opportunistic methods}, which bootstrapped off actual results from executed queries and views~\cite{bruno2002exploiting,bruno2001stholes}.  In a similar spirit, workload results have also been used to make adjustments to correlated predicates~\cite{markl2003leo}.

In order to adapt in a more principled fashion, \emph{learned database systems} replace traditional database components (\textit{e.g.,} the cardinality estimator and underlying summary structures, query optimizer, and more) with ML models (\textit{e.g.,} MSCN~\cite{kipf2019learned}, Tree CNN~\cite{marcus2019plan}, HAL~\cite{ma-active-learning}, and more), in the hope of improving the performance of database components as the system learns to better estimate performance from past queries. Among existing proposals for learned database systems, most are query-driven (also known as workload-driven), which means they are trained from a query workload.  Indeed, this paper focuses on such a setting. We can formalize the notion of a query-driven, learned database system and apply it to various components, such as cardinality estimators, as follows.

\begin{definition} [Query-Driven Learned Database Systems]
  Let $\mathcal{CP}$ be a database component we want to optimize (\textit{e.g.,} cardinality estimators). In the scenario of learned database systems, a record (log) of previously executed workload $\mathcal{W}$, which consists of a set of $n$ queries $\{q_i\}_{i=1}^n$ and their outcome or label $\{o_i\}_{i=1}^n$ (\textit{e.g.,} cardinality) is available. Query-driven learned database systems aim to train a machine learning model $\mathcal{M}$ to make predictions for $\mathcal{CP}$, by leveraging the historical workload $\mathcal{W}$.
\end{definition}

A shortcoming of query-driven learned database systems is that they learn solely from the characteristics of the training workload, and do not leverage any of the \emph{domain knowledge} that more traditional DBMSs use as the basis of cost estimation.

\subsection{Database Domain Knowledge} \label{subsection.pre.knowledge}

In traditional query optimization, cardinality estimation relies on a combination of knowledge derived from the database schema, statistical information about the database instance (from histograms, sketches, indices, etc.), and configuration information --- all incorporated into estimation formulas developed based on the domain.



\subsubsection{\textbf{Definitions and notation}}

Database domain knowledge refers to a wide range of \emph{rules} and  \emph{principles} related to the design, performance, and management of database systems, augmented by information about database schemas and constraints. For example, domain knowledge for cardinality estimation might include the knowledge that a $1:m$ join on a primary key-foreign key will result in $m$ results; that a grouping attribute has unique values; or that a specific functional or inclusion dependency holds. Domain knowledge for cost estimation might include the knowledge that the cost of a nested loop join grows quadratically.

Although domain knowledge can be incorporated in a variety of contexts, in order to limit our scope, in this paper we address the problem of cardinality estimation (which is a prerequisite for effective cost estimation).  We develop a general methodology for incorporating knowledge in the form of constraints, and we focus on two families of constraints:  primary key-foreign key (representing the most common ways functional and inclusion dependencies are captured in schemas), and query containment (specifically, based on range conditions), since they are constraints that we have access to in a DBMS and have high utility in cardinality estimation. For a query $q$, we define two \emph{interchangeable} notations, $c(q)$ and $|q|$, to be the cardinality of $q$.

\begin{definition} [Primary Key (PK), Foreign Key (FK)] A Primary Key (PK) is a column (or set of columns) in a table that uniquely identifies each row in that table, i.e., functionally determines all attributes in that column. A Foreign Key (FK) is a column (or set of columns) in one table that joins with the PK in another table in a dependency-preserving fashion.
\end{definition}



\begin{definition} [Query Containment and Complementarity Properties]
Let $q_1$ and $q_3$ be two union-compatible queries over schema $\Sigma$, such that $q_1(\cdot) \subseteq q_3(\cdot)$ for any instance of $\Sigma$.  We refer to this as a case of \emph{query containment}~\cite{abiteboul1995foundations}. The cardinality of $q_1(\cdot)$ will always be less than or equal to the cardinality of $q_3(\cdot)$.

Define a query $q_2(\cdot) = q_3(\cdot) - q_1(\cdot)$ to be the \emph{complement} of $q_1$ with respect to $q_3$.  Directly reasoning from the properties of relational difference, we know that the cardinality of $q_3(\cdot)$ must be equal to the sum of the cardinalities of $q_1(\cdot)$ and $q_2(\cdot)$.
\end{definition}


\subsubsection{\textbf{Representing domain knowledge as constraints}}
We leverage the concept of \emph{constraints} to encode domain knowledge within query-driven learned databases. More specifically, we target three types of constraints.
\\

\noindent \textit{Constraint Class 1 (PK-FK Inequality Constraint).} Consider a query $q$ that joins the table $A$ (with PK) with another table $B$ (with FK) through the PK-FK join, \textit{i.e.,} $A \bowtie B$. Note that here $B$ can be the join result of multiple tables.
Then, suppose the subqueries on tables $A, B$ are $q_A, q_B$, respectively. From the definition of PK-FK join, one can easily derive the inequality constraint: \underline{$c(q) \leq c(q_B)$}, since each row in $B$ can only be joined with at most 1 row in $A$. 
\\

\noindent \textit{Constraint Class 2 (PK-FK Equality Constraint).}
This type of PK-FK constraint is a fine-grained version of Constraint Class 1. If there is no predicate on $A$, we can reduce the inequality constraint to an exact equality constraint: \underline{$c(q) = c(q_B)$} since now we can guarantee each row in $B$ can join \emph{exactly} 1 row in $A$.
\\

\noindent \textit{Constraint Class 3 (Consistency Constraint).}
The consistency constraint is derived from the query containment and complementarity properties. We define the consistency constraint over three queries: the cardinality of a range query $q$ is equal to (or larger than, depending on the \texttt{NULL} ratio in the queries column\footnote{In this paper, we will use the consistency constraint as $c(q_1) = c(q_2) + c(q_3)$ because we observe in our experiments that \texttt{NULL} values have very little effect in the cardinality results, due to their very small ratios in all columns.}) the sum of two range queries ($q_1$, $q_2$) that are split from the original query, \textit{i.e.,} \underline{$c(q_1) = c(q_2) + c(q_3)$}. Note that the separated queries have the same join graph as the original query, and they do not overlap in query ranges.

\subsection{Does Query-Driven Learned Databases Learn Domain Knowledge?}\label{section.motivation.exp}

In this subsection, we meticulously devise and execute a set of experiments to evaluate if learned databases can acquire domain knowledge from training queries. We use the well-known MSCN~\cite{kipf2019learned} model for the task of learned cardinality estimation, targeting two domain knowledge constraints introduced in Section~\ref{subsection.pre.knowledge} (\textit{e.g., consistency constraint and PK-FK Inequality constraint}).

\noindent \textbf{Experimental Details.} We trained the MSCN model of the default model architecture in~\cite{mscncode} with a large set (20K) of training queries on the DSB~\cite{ding2021dsb} dataset. 


For testing the consistency constraint, we randomly generate query triples that match the consistency constraint property and evaluate their predictions. We measure the ratio of model predictions $r_{cc} = |\hat q_1| / (|\hat q_2|+|\hat q_3|)$. We also randomly generate query pairs that match the PK-FK inequality constraint property and calculate the ratio of model predictions $r_{pk-fk} = |\hat q_1| / |\hat q_2|$. Please see Section~\ref{section.exp.setup} for details on generating queries for constraints.

\noindent \textbf{Takeaways.}  We plot the distribution of the (log of) ratio for each constraint in Figure~\ref{fig.example}.  For the consistency constraints in the left subfigure, a ratio whose log is closer to 0 indicates better performance. For the right subfigure indicating the PK-FK inequality constraint, the left side of the red dashed line is possibly correct (satisfying the inequality), whereas the right side is definitely incorrect. From the figure, we observe two points:

\begin{itemize}  [leftmargin=*]
    \item \textbf{Learned query-driven cardinality can be inconsistent.} We define a significant violation if the ratio (without log) $r_{cc} > 2$ or  $r_{cc} < {1}/{2}$. We observe $22\%$ cases of significant violation of the consistency constraint. Moreover, the minimal and maximum of $r_{cc}$ are $2.7 \cdot 10^{-5}$ and 929, respectively, indicating MSCN can \emph{largely} violate the consistency constraint.

    \item \textbf{Learned query-driven cardinality does not well capture PK-FK constraint.} We observe $10\%$ cases where the ratios $r_{pk-fk}$ (without log) are a clear violation of the PK-FK inequality constraint, \textit{i.e.,} $r_{pk-fk} > 1$. The maximum  $r_{pk-fk}$ is even 31, which deviates considerably from possible correct values ($\leq 1$).

\end{itemize}

The lack of domain knowledge in query-driven learned models underscores the need for a comprehensive solution to integrating the knowledge into the model, as domain knowledge represents fundamental rules and principles that ought to be encoded to a (learned) database.

\section{Learning Domain Knowledge with \name}
\label{sec:domain-knowledge}


\begin{figure}
\centering
\includegraphics[width=0.47\textwidth]{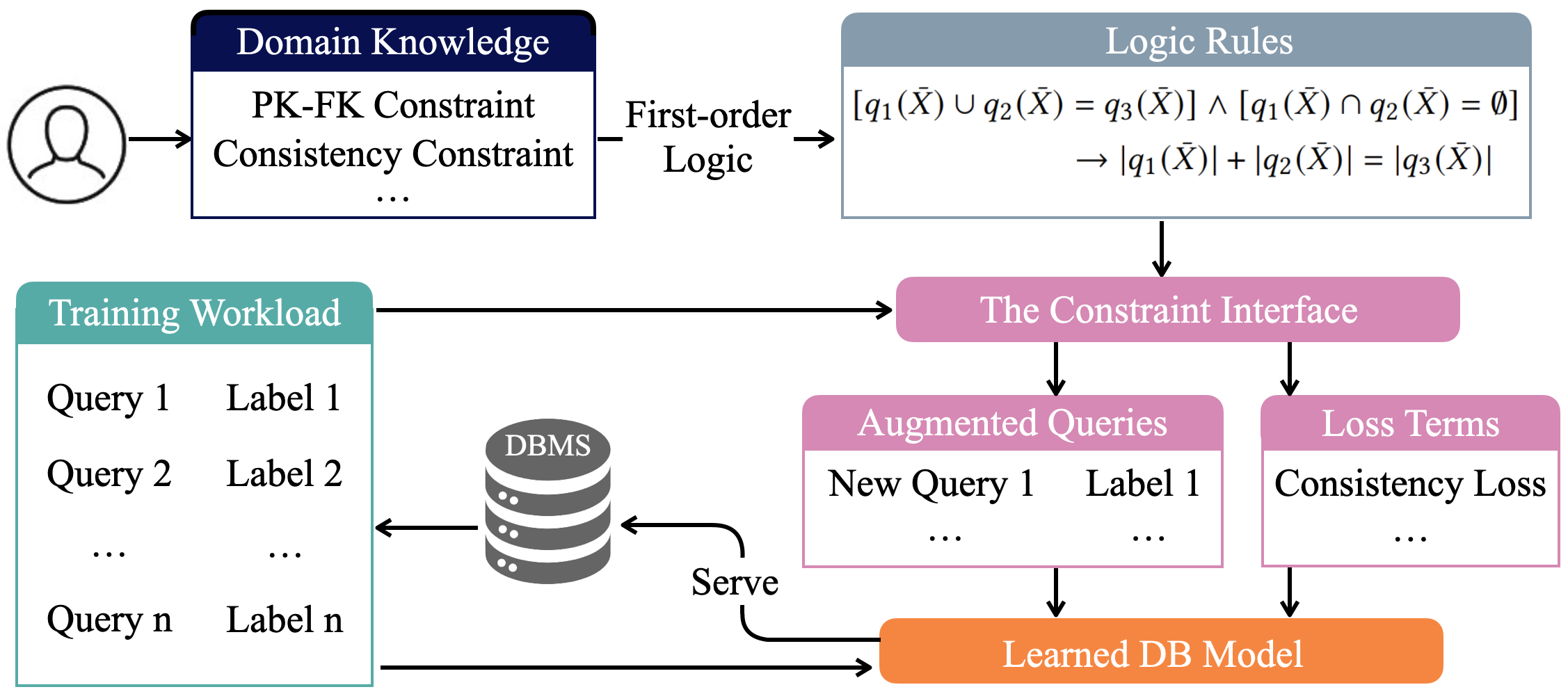}
\vspace{-0.7em}
\caption{Overview of \name.}\label{fig:overview}
\end{figure}

\name mostly follows the standard workload-driven learned cardinality estimation model~\cite{kipf2019learned, hasan2020deep} in which a user-provided workload with queries and true cardinalities is provided, and a model is trained using this data. Instead of training exclusively on a user-provided training workload, \name{} \emph{augments the training workload} using information gathered from logical constraints. Logical constraints are used to augment the training workload in two ways: first, and most similar to traditional data augmentation for machine learning tasks, additional, provably-correct training samples are added to the training set. Second, \name{} adds additional terms to the model's loss function. Each additional term represents a logical constraint, in that the additional term is minimized when the model's prediction obeys the constraint. 

Figure~\ref{fig:overview} illustrates the basic workflow: first, as in any workload-driven learned database, we start with a sample workload from the user. This sample workload is used to generate the \emph{empirical training} loss, which can be any standard regression loss function. Second, various pieces of domain knowledge (e.g., PK-FK constraints) are used to generate (a) augmented, provably-correct training samples, which are added to the training workload and thus to the empirical training loss, and (b) a \emph{logic loss} function which represents a differentiable term that is minimized when the learned model's predictions do not violate a constraint imposed by domain knowledge. Third, a learned model is trained by minimizing the sum of the empirical training loss and the logic loss. Finally, the newly trained model is used to serve queries from the DBMS.

This process is formalized in Algorithm~\ref{alg:alg1}, and expanded upon in each of the following sections. We initially (in Section~\ref{sec:constraints}) focus on the heart of the algorithm, detecting which \emph{constraints} are applicable to a query sampled from the workload: we describe each piece of domain knowledge that \name{} translates into augmented training samples and logic loss terms. Section~\ref{sec:con_consistency} shows how consistency constraints can be encoded in loss functions, Section~\ref{sec:con_joins} shows how primary key joins can be used to generate additional training samples, and Section~\ref{sec:con_ineq_joins} shows a more complex case where modifying the predicates of a join query can produce an upper-bound on the resulting cardinality. In Section~\ref{sec:training_loop}, we explain how the entire system is trained end-to-end, with a detailed discussion of the algorithm.

\begin{figure}[tb]
\begin{algo}[caption={Main training loop}, label={alg:alg1}]
# lossFn starts initialized to default
foreach epoch:
    while executionWorkload has not been fully sampled:
      minibatch = sampleWOReplacement(executionWorkload)
      foreach (query, label) in minibatch:
        choose constraint from applicableConstraints(query):
          (lossFn, additionalTraining) = 
            applyConstraint(constraint, query, label, lossFn)
          minibatch.append(additionalTraining)
    
          applySGD(minibatch, lossFn)
\end{algo}
\end{figure}

\subsection{Constraints}
\label{sec:constraints}

In this section, we describe \name{}'s \emph{constraints}, a library of domain knowledge about query cardinality. We first explain the common interface that each constraint applies. Next, we give three examples of logical constraints. We explain how this library of constraints is used to train a cardinality estimation model in Section~\ref{sec:training_loop}.

\paragraph{The constraint interface} \name{} represents constraints as two functions: (1) \texttt{applicable}, which determines if a specific constraint applies to a given query, and (2) \texttt{augment}, which generates either an additional piece of training data or an additional loss term from the given applicable query.

Formally, for a given query $q_i \in Q$, we define a \emph{constraint} $C$ as a pair of functions:
\begin{enumerate}
    \item{$\texttt{applicable}: Q \to \{0, 1\}$, which maps a query to either zero (false), indicating that the constraint cannot be used to augment that query, or one (true), indicating that the constraint can augment the query.}
    \item{$\texttt{augment}: Q \to (Q, \mathbb{N})^n \times F$, which maps an applicable query to $n$ additional training instances (with $n$ possibly equal to zero), and a differentiable function $F$ that is minimized when the constraint is satisfied ($F$ may be zero).}
\end{enumerate}

To illustrate these two functions, we next give three examples of constraints used by \name: 1) \emph{consistency constraints} based on containment and complementarity, 2) PK-FK \emph{equality} constraints based on joins with primary keys, and 3) PK-FK \emph{inequality} constraints also based on joins between primary keys and foreign keys, but with different criteria. For each constraint, we will first formally define the constraint in first-order logic, then explain the \emph{applicable} and \emph{augment} functions.

\subsubsection{\textbf{Learning Consistency Constraints}}
\label{sec:con_consistency}

The \emph{consistency constraint} is based on our domain knowledge about the query containment and complementarity properties. We first formally define the constraint and its applicability. We then give an example of how this constraint could be applied.

\paragraph{Constraint.}
Formally, let $q_1(\bar X), q_2(\bar X), q_3(\bar X)$ be queries over schema $\Sigma$, returning attributes $\bar X$. Then, the following first-order logic assertion holds, by property of the relational algebra:
\begin{equation}
\label{eq:consistency}
    \begin{aligned}
        [q_1(\bar X) \cup q_2(\bar X) &= q_3(\bar X)] \land [q_1(\bar X) \cap q_2(\bar X) =  \emptyset] \\
        & \rightarrow  |q_1(\bar X)| + |q_2(\bar X)| = |q_3(\bar X)|
   \end{aligned}
\end{equation}


\paragraph{Applicability.}
In our implementation, this constraint can be applied to any query where there exists at least one column $c$ not returned in our query $q_3$, which has multiple values in its active domain.\footnote{The existence of the additional column is not strictly necessary, but makes it easier to define the query subsets without affecting the predicates in the original query.}

\paragraph{Loss.}
Given a training query $q_3$, we generate queries $q_1$ and $q_2$. To do so, we first select a column $C$ that is not read by $q_3$. We randomly sample a \emph{split point} $p$ from $C$. We then generate $q_1$ as $q_1 = \sigma_{c < p}(q_3)$, and we generate $q_2$ as $q_2 = \sigma_{c \geq p}(q_3)$. While we do not know the cardinality of $q_1$ or $q_2$, we know by construction that $|q_1| + |q_2| = |q_3|$. We add a term to the loss function to encourage the trained model to respect this constraint. Let $\hat{q}$ be the model prediction of the cardinality of query $q$. We then add the term $qerror(|\hat{q_3}|, |\hat{q_1}| + |\hat{q_2}|)$ to the loss function:


\begin{equation}
\label{eq:qerror_consistency}
   qerror(|\hat{q_3}|, |\hat{q_1}| + |\hat{q_2}|) = \max\left(\frac{|\hat{q_3}|}{|\hat{q_1}|+|\hat{q_2}|}, \frac{|\hat{q_1}|+|\hat{q_2}|}{|\hat{q_1}|}\right)
\end{equation}

Note that Equation~\ref{eq:qerror_consistency} is minimized precisely when the predicted cardinality of each query satisfies the constraint in Equation~\ref{eq:consistency}. Note that any regression loss function -- such as mean-squared error or mean-absolute error -- would also satisfy this condition, but we use the symmetric relative error (Q-error) to be consistent with prior work~\cite{kipf2019learned, deepdb, negi2023robust}.
Interestingly, Equation~\ref{eq:qerror_consistency} does not depend on the true cardinality of $q_3$, $|q_3|$, which is known ahead of time. While in practice, substituting the known true cardinality $|q_3|$ for $|\hat{q_3}|$ would likely improve training performance, it is not strictly necessary.


\paragraph{Data augmentation.}
For this constraint, we choose not to add additional training instances, since we effectively are adding information about $q_1, q_2$ into the loss for $q_3$, as described above.

\begin{example}
    Consider a relation with two real-valued attributes $R = (A \in \mathbb{R}, B \in \mathbb{R})$ and a query $q_3 = \sigma_{A > 0.5} (R)$. Since $q_3$ does not read attribute $B$, we can rewrite $q_3$ to:
    \begin{equation*}
        q_3 = \sigma_{A > 0.5} (R) = \sigma_{A > 0.5}(\sigma_{B < 0.25}(R)) \cup \sigma_{A > 0.5}(\sigma_{B \geq 0.25}(R))
    \end{equation*}

    While this rewrite obviously has no performance benefits, we can infer the following cardinality constraint from it:
    \begin{equation*}
        |q_3| = |\sigma_{A > 0.5}(\sigma_{B < 0.25}(R))| +  |\sigma_{A > 0.5}(\sigma_{B \geq 0.25}(R))|
    \end{equation*}

    Letting $q_1 = \sigma_{A > 0.5}(\sigma_{B < 0.25}(R))$ and $q_2 = \sigma_{A > 0.5}(\sigma_{B \geq 0.25}(R))$, we can use Equation~\ref{eq:qerror_consistency} to build a loss term that will be minimized when the model has a consistent prediction for the size of $q_1$ and $q_2$. That is, the loss term is minimized when the model predicts $|\hat{q_1}|$ and $|\hat{q_2}|$ such that $|\hat{q_1}| + |\hat{q_2}| = |q_3|$.
\end{example}


\subsubsection{\textbf{Learning Equality Constraints from Joins}}
\label{sec:con_joins}
The next constraint considers how to \emph{augment} a query by joining it on a foreign key-primary key association, in a lossless fashion. The join on a primary key should not affect the cardinality of the result. We first formally define the constraint and then give an example.

\paragraph{Constraint.}
Let $T$ be a relation in schema $\Sigma$, $q_1, q_2$ be queries over $\Sigma$, and $\bar X \rightarrow T$ be a functional dependency over $\Sigma$. Then:
\begin{equation}
    \begin{aligned}
        T(\bar X, \bar Y) & \wedge q_1(\bar X, \bar Z) = q_2(\bar X' \subseteq \bar X, \bar Y, \bar Z) \land (FD: \bar X \rightarrow T) \\
       &  \rightarrow  |q_1(\bar X, \bar Z)| = |q_2(\bar X')|
   \end{aligned}
\end{equation}

\paragraph{Applicability.}
This constraint can only be applied to a query $q_1(\bar X, \bar Z)$, if we can find a table $T$ not referenced in $q_1$, yet joinable with $q_1(\bar X, \bar Z)$ on its primary key. In other words, we must be able to define $q_2(\bar X' \subseteq \bar X, \bar Y, \Bar Z) = q_1(\bar X, \bar Z) \wedge T(\bar X, \bar Y)$, where $\bar X$ is a key for $T$.

\begin{figure}[tb]
\centering
\includegraphics[width=0.47\textwidth]{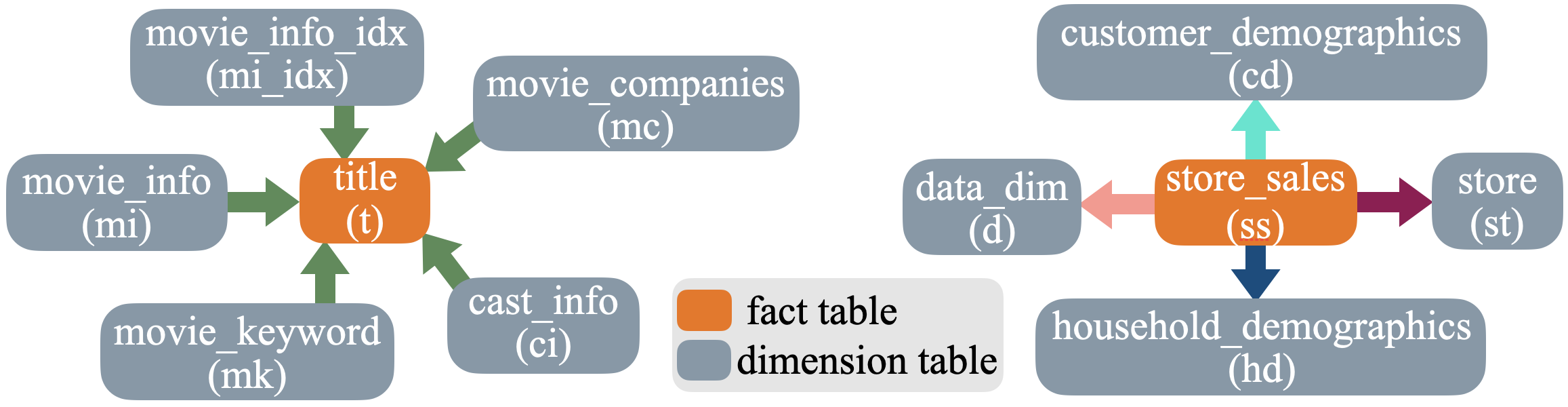}
\caption{Sub-schemas for datasets. Left: IMDb. Right: DSB}\label{fig.join}
\end{figure}

\paragraph{Loss.} We do not add any additional loss terms for this constraint, but instead generate new training samples.


\paragraph{Data augmentation.}
We add $q_2(\bar X')$ to our training set, with label $|q_1|$ based on the equality constraint.

\begin{example}
Figure~\ref{fig.join} shows a simplified schema from the IMDb~\cite{leis2015good} and DSB~\cite{ding2021dsb} benchmarks. Both datasets have a \textcolor[HTML]{e2792e}{fact table} and several \textcolor[HTML]{8898a6}{dimension tables}. 
In the IMDb schema, the fact table \texttt{title} has a primary key, and all dimension tables have a foreign key to that primary key. Consider a training query that does not contain the fact table, such as $q_1 = \sigma_{P}(\mbox{\texttt{mc}})$. Next, let $q_2$ be the join of $q_1$ with the fact table: $q_2 = q_1 \bowtie \mbox{\texttt{t}} = \sigma_{P}(\mbox{\texttt{mc}}) \bowtie \mbox{\texttt{t}}$. Regardless of what the predicate $P$ does, the join with \texttt{title} cannot add or remove any rows from $q_1$ (since the join is a PK-FK join). Thus, we have the constraint $|q_1| = |q_2|$, and we can add $q_2$ to our training set with known cardinality equal to $q_1$.

\end{example}

\subsubsection{\textbf{Learning Inequality Constraints from Joins}} This constraint considers queries that join a relation on that relation's primary key. We first formally define the relation and then give an example.

\label{sec:con_ineq_joins}

\paragraph{Constraint.}
Consider a query $q_2$ that includes a join with some table $T$ on $T$'s primary key, and also contains a predicate $\phi(\cdot)$ over the attributes of $T$. Removing the join with $T$ and any predicates on $T$ results in a less-constrained query $q_1$, which must have at least as many results as $q_2$.
%
\\
\begin{equation}
\label{eq:join_ineq}
    \begin{aligned}
        T(\bar X, \bar Y) & \wedge \phi(\bar X, \bar Y) \wedge q_1(\bar X, \bar Z) = q_2(\bar X' \subseteq \bar X, \bar Y, \bar Z) \land (FD: \bar X \rightarrow T) \\
       &  \rightarrow  |q_1(\bar X, \bar Z)| \geq |q_2(\bar X')|
   \end{aligned}
\end{equation}

\paragraph{Applicability.}
This constraint can only be applied to a query $q_2(\bar X, \bar Z)$, if it includes a join with a table $T$ on $T$'s primary key.

\paragraph{Loss.} Unlike in the previous two constraints, here we do not have an equality predicate. Instead, we have an inequality: any model that predicts a $|\hat{q_1}|$ and a $|\hat{q_2}|$ such that $|\hat{q_1}| \geq |\hat{q_2}|$ is potentially correct, but any model that predicts a $|\hat{q_1}|$ and a $|\hat{q_2}|$ such that $|\hat{q_1}| < |\hat{q_2}|$ is definitely wrong. Thus, we want to penalize models for any incorrect prediction, but not penalize models for potentially correct predictions. In terms of a loss function, and in contrast with the previous two constraints, we seek asymmetry: \emph{any} prediction of $|\hat{q_1}|$ that is greater than or equal to $|\hat{q_2}|$ should receive no penalty, but a prediction of $|\hat{q_1}|$ that is less than $|\hat{q_2}|$ should be penalized. 

To accomplish this, we use the notion of \textit{counterexample-guided learning}~\cite{sivaraman2020counterexample} to define the loss function for inequality constraints. The high-level idea is to create a term that grows when the inequality constraint is violated, but is zero whenever the inequality constraint is not violated. 
Such a function $f$ can be defined in a simple piecewise fashion:
\begin{equation*}
    f(|\hat{q_1}|, |\hat{q_2}|) = \begin{cases} 
      0 & |\hat{q_2}| - |\hat{q_1}| \leq 0 \\
      |\hat{q_2}| - |\hat{q_1}| & |\hat{q_2}| - |\hat{q_1}| > 0 \\
   \end{cases}
\end{equation*}

Conveniently, we can define $f$ in terms of the ReLU activation function~\cite{agarap2018deep}, which has efficient implementations in modern deep learning libraries: $f(\hat{q_1}, \hat{q_2}) = ReLU(\hat{q_2} - \hat{q_1})$. Alternatively, we can also replace the $|\hat{q_2}| - |\hat{q_1}|$ with $qerror(\hat{q_2}, \hat{q_1})$ if $|\hat{q_2}| - |\hat{q_1}| > 0$. We call the definition of $f$ a labeling-by-bound strategy since we basically label $|\hat{q_1}|$ with its lower bound $|\hat{q_2}|$. We will revisit this strategy in Section~\ref{section.plabel}.

\paragraph{Data augmentation.}
For this case, we do not add additional training instances, since we have no ground-truth labels on $q_1$. 

\begin{example}
Consider the DSB schema in Figure~\ref{fig.join} and a query $q_2$, in which the fact table \texttt{ss} joins with dimension table \texttt{st} and dimension table \texttt{hd}: $q_2 = \mbox{\texttt{ss}} \bowtie \mbox{\texttt{st}} \bowtie \sigma_P(\mbox{\texttt{hd}})$. Note that the fact table \texttt{ss} joins with \texttt{st} and \texttt{hd} on \texttt{st}'s and \texttt{hd}'s primary keys. Thus, we can create $q_1$ by dropping the join with \texttt{hd}: $q_1 = \mbox{\texttt{ss}} \bowtie \mbox{\texttt{st}}$ (we also could have dropped the join with \texttt{ss}). By Equation~\ref{eq:join_ineq}, we have that $|q_1| \geq |q_2|$. Intuitively, this is because the join with $\sigma_P(\mbox{\texttt{hd}})$ in $q_2$ can only produce the same or fewer rows, since we are joining with the primary key of \texttt{hd}. We can enforce this constraint by minimizing $f(\hat{q_1}, \hat{q_2}) = ReLU(\hat{q_2} - \hat{q_1})$, since $f(\hat{q_1}, \hat{q_2})$ will be zero whenever the constraint is satisfied, but positive otherwise.
\end{example}

\subsection{\name{}'s Training Loop}
\label{sec:training_loop}

\name{}'s training loop closely matches the training loop of a standard deep learning pipeline, using stochastic gradient descent to minimize a loss function. The training loop is described in pseudo-code in Algorithm~\ref{alg:alg1}. \name{} trains its neural network model in \emph{epochs}, where each epoch represents a pass over the training data. Each epoch is composed of a sequence of \emph{minibatches}, where each minibatch contains a small sample (without replacement) from the training set. The current network weights are used to make a prediction for the samples in the minibatch, which are then numerically compared against the ground truth data using a loss function. The derivative of this loss function is used to optimize the network weights. An epoch ends when there is no more data left to sample, and a new epoch begins.

Unlike the standard deep learning pipeline, \name{} augments each minibatch and the loss function during each iteration. For each query and label in the minibatch, \name{} first generates a set of \emph{applicable} logical constraints. Then, a subset of applicable logical constraints is chosen (see Section~\ref{sec:constraints}). Each of the chosen logical constraints then produces one or both of (1) additional training data, which is added to the current minibatch, and/or (2) additional loss terms, which are added to the current loss function. We will discuss how to choose the subset of applicable logical constraints in Section~\ref{section.multupleconstraints}.

After adding additional samples to the minibatch and adding additional terms to the loss function, stochastic gradient descent is used to minimize the combined loss function. Formally, let $T$ be the set of training queries in the minibatch and let $L$ be the set of additional loss terms generated from constraints. Then we minimize:
\begin{equation*}
\sum_{q_i \in T} qerror(q_i, |q_i|) + \omega \sum_{loss \in L} loss()
\end{equation*}

\noindent where $\omega \geq 0$ is a hyperparameter that controls the balance between correctly predicting the cardinality of the training queries (most extreme at $\omega = 0$) and respecting the generated constraints (more extreme as $\omega$ increases). Note that $\omega$ can be thought of as a Lagrange multiplier: regardless of the value of $\omega$, the function has the same \emph{global} minima (this is true because there exists some perfect set of cardinality estimates). Of course, in practice, gradient descent is not likely to find such global minima, so tuning may be required.
\section{Optimizing \name}\label{section.qsampling}
In this section, we optimize our system in terms of the labeling strategy for inequality constraints and handling multiple constraints.

\subsection{Pseudo-Labeling for Inequality Constraints} \label{section.plabel}
While simple and efficient, the labeling-by-bound strategy might not be accurate if the true label is far from the bound. To mitigate this issue, we propose an alternative labeling method for inequality constraints, inspired by the concept of Pseudo-Labeling~\cite{berthelot2019mixmatch} in semi-supervised learning.

\smallskip
\noindent \textbf{Pseudo-Labeling.} Pseudo-labeling aims to provide pseudo-labels for unlabeled data based on the predictions of the model trained on labeled data. To do so, for an unlabeled data point, it computes the average of the model’s predicted class distributions across all the K augmentations as the final "guess" for the unlabeled point. This is based on the assumption that \emph{small perturbations (e.g., rotation to an image) do not change its class}. Therefore, one can use the "neighbors" (perturbations) of an unlabeled point to approximate its label.

\smallskip
\noindent \textbf{Applying Pseudo-Labeling.} To apply pseudo-labeling to label generated queries $q'$, the key point is to identify an "invariant" characteristic, ensuring that the transformation into a query does not alter its cardinality. Fortunately, the consistency constraint is a natural fit for this purpose. Specifically, for each new query $q'$ that violates an inequality constraint, we 1) randomly generate $k$ sets of split query pairs $\{q_{i1}, q_{i2},\}_{i=1}^k$ from $q'$ using consistency constraint, and 2) compute the mean value of the sum of the model predictions for queries in each pair as the "guess" label for $q'$.

    


\subsection{Efficient Training with Multiple Constraints} \label{section.multupleconstraints}
Readers may notice that there could be multiple constraints applicable to a single query. In fact, most queries in our experiments are eligible for both the consistency constraint and PK-FK constraint. And even for the same constraint such as the PK-FK constraint, there could exist multiple ways to incorporate the constraint.
In this subsection, we proceed to introduce how \name achieves efficient training of a single model with multiple constraints.

\smallskip
\noindent \textbf{First Attempt.}  One could apply \emph{all} applicable constraints to every training query and add the resulting logic loss terms to the training loss. While effective, this strategy will substantially increase the training complexity and computational cost. 

\smallskip
\noindent \textbf{Practical Multiple Constraint Training.}  We propose to use \emph{random choice} to achieve the goal of reducing the training overhead while not sacrificing the model performance. As was illustrated in Algorithm~\ref{alg:alg1}, for each training query, \name randomly chooses an applicable constraint (\textit{e.g.,} PK-FK constraint) and applies its corresponding action (augmenting the data and/or loss function).


\smallskip
\noindent \textbf{Future Extension.}  Although we empirically observe the simple random choice-based training strategy is efficient and effective in enabling a single model to learn multiple constraints, an interesting extension would be to explore the potential of using multi-armed bandit algorithms~\cite{marcus2022bao}  and reinforcement learning techniques~\cite{kaelbling1996reinforcement} to choose the best constraint and action at each step.


\section{Experimental Evaluation}\label{section.evaluation}
In this section, we implement \name with the well-known query-driven cardinality estimation model, MSCN. We aim to answer the following research questions.
\begin{enumerate} 
\item Does \name reduce constraint violations in MSCN predicitons?
\item Can \name improve cardinality estimation for MSCN?
\item Can \name help query-driven MSCN work on fewer training queries?
\item Does \name help query-driven MSCN achieve a better query end-to-end performance? 
\item What are the behaviors of optimization strategies in \name?

\end{enumerate}

\subsection{Experimental Setup} 
\label{section.exp.setup}

\noindent \textbf{Datasets.} As introduced in Section~\ref{sec:domain-knowledge}, we conduct all experiments on two datasets: IMDb~\cite{leis2015good} and DSB~\cite{ding2021dsb}. IMDb is a complex real-world dataset on films and television programs from the Internet Movie Database.  DSB is an extension to the TPC-DS benchmark\cite{poess2002tpc}, which uses more complex data distributions and challenging query templates. We
populate a 50GB DSB database with the default physical design
configuration. We focus on the star-join schema (Figure~\ref{fig.join}) which constrains 6 and 5 relations for IMDb and DSB, respectively.

\smallskip
\noindent \textbf{Training Workloads.} For the IMDb dataset, we leverage the training queries from~\cite{kim2022learned} which contain up to 5 joins, as the workload to train the MSCN model. We discard all queries that contain a negation operator  (very rare in common benchmarks) and use 30K queries to train the model.
Note that we do not use the original training set from~\cite{mscncode}, which only contains queries of up to 2 joins. In fact, MSCN trained on the set of queries of limited join types~\cite{han2021CEbenchmark} is proven to perform \emph{much worse} than MSCN trained on queries of diversified join types (up to 5 joins) ~\cite{kim2022learned}, in terms of both prediction accuracy and query running time performance. We observed this fact in our experiments as well. 


For the DSB dataset, we generate 20K training queries following the join schema of Figure~\ref{fig.join}. From the lessons from the comparison of the two MSCN workloads, we generate DSB training queries following the two principles: (1) \emph{coverage}: we want to cover as many join graphs as possible, and (2) \emph{diversity}: we want to diversify the number of predicates and operators. Specifically, we first sample the number of joins from $[0, 4]$, and then randomly choose a join graph of the sampled join number. We then sample the number of predicates from $[1, 8]$ and the queried columns in the join graph accordingly. For numerical columns, we sample the operators from $\{<, \leq, =, \geq, > \}$ at random and sample a value from the active domain. For categorical columns, we only use the equality filter ($=$).


\smallskip
\noindent \textbf{Training Details.} We train the MSCN model on an Amazon SageMaker \textbf{ml.g4dn.xlarge} node, and conduct end-to-end performance experiments on an EC2 \textbf{r5d.2xlarge} node for the IMDb dataset and on an EC2 \textbf{c5.9xlarge} node for the DSB dataset. For \name, we turn on the pseudo-labeling mechanism (number of "perturbations" $k=5$) for inequality constraints and the random choice-based efficient training strategy.

\smallskip
\noindent  \textbf{Test Queries.} 
In the machine learning literature, it is widely acknowledged that models tend to exhibit superior performance on training data, or data within the same distribution (which we abbreviate in this paper as \textbf{In-Dis}(tribution) data) compared to out-of-distribution (\textbf{OOD}) test data.  Therefore, for estimation accuracy experiments, we consider two kinds of test workloads for both datasets. For \emph{In-Dis Queries}: we randomly sample 2K test queries from the same distribution of the training workloads. Note that the training and test queries do not overlap. For \emph{OOD queries}: we sample test queries from a different query distribution from the training workloads. For IMDb, we directly use all queries and subqueries of the JOB-light benchmark. For DSB, we randomly sample 500 queries from the In-Dis test queries and generate all subqueries (3984 in total) as the OOD queries. We apply the widely used metric of q-error~\cite{moerkotte2009preventing}. We also include the performance of PostgreSQL for reference.

\subsection{Can \name Learn to Predict the Effects of Database Constraints?} \label{section.exp.vio}
We begin by validating that \name can be used to predict cardinalities taking into account the effects of constraints.  We assess this by measuring how widely often the predicted cardinalities violate the effects of known constraints.
 
\begin{table}[ht]
\centering
\scalebox{1}{
\begin{tabular}{|c|c|c|c|}
  \hline
  \backslashbox{Constraint}{Model}  &MSCN &  \makecell{Constrained \\ MSCN}  &  \makecell{Constrained \\ MSCN-50\%}  \\
  \hline
{Consistency}  & $3\%$ &  $1.2\%$ ($\downarrow60\%$) & $1.8\%$ ($\downarrow40\%$)\\
  \hline
  {PK-FK Inequality }  & $8.9\%$ & $4.7\%$ ($\downarrow47\%$) & $5.1\%$ ($\downarrow46\%$)\\ 
  \hline
  {PK-FK Equality } & $4.0\%$ & $1.6\%$ ($\downarrow60\%$)& $2.3\%$ ($\downarrow43\%$)\\
  \hline
\end{tabular}}
\caption{\mbox{Ratio of constraint violations on IMDb ($\downarrow$ \% reduction)}}
\label{exp.table.numvio.imdb}
\end{table}

\begin{table}[ht]
\centering
\scalebox{1}{
\begin{tabular}{|c|c|c|c|}
  \hline
  \backslashbox{Constraint}{Model}  &MSCN &  \makecell{Constrained \\ MSCN}   &  \makecell{Constrained \\ MSCN-50\%}  \\
  \hline
{Consistency}  &  $7.0\%$ & $2.4\%$ ($\downarrow64\%$) & $4.2\%$ ($\downarrow40\%$)  \\
  \hline
   {PK-FK Inequality}  & $10\%$ & $4.9\%$ ($\downarrow51\%$)  & $5.4\%$ ($\downarrow46\%$) \\ 
  \hline
   {PK-FK Equality} & $3.5\%$  & $1.5\%$ ($\downarrow57\%$) & $2.0\%$ ($\downarrow43\%$)  \\
  \hline
\end{tabular}}
\caption{\mbox{Ratio of constraint violations on DSB ($\downarrow$ \% reduction)}}
\label{exp.table.numvio.dsb}
\end{table}

We leverage the test queries from Section~\ref{section.exp.card} to generate new queries according to each constraint applied (by following the method for constraint-guided query generation described in Section~\ref{sec:constraints}). We evaluate the three constraints on both datasets and present the results (the ratio of "significant" constraint violations) in Tables~\ref{exp.table.numvio.imdb} and~\ref{exp.table.numvio.dsb} for IMDb and DSB, respectively. We count "significant" violations, determined as follows:

\begin{itemize} [leftmargin=*]  
\item For equality constraints, the model predicts cardinality for the right side that is \textbf{2 times} larger or smaller than the prediction for the left side.
\item For inequality constraints, when the predicted cardinalities violate the inequality.
\end{itemize}

\noindent
(Intuitively, it is harder to learn an exact equality condition than an inequality condition since the valid output range for an inequality constraint is much larger than that for an equality constraint, which is why we build a slack factor into the first case.)


We observe the following key points. First, it is not unusual for query-driven models' predictions to violate basic constraints. For example, the ratio of violations is $8.9\%$ for the PK-FK Inequality constraint on IMDb and is $10\%$ for the PK-FK Inequality constraint on DSB. Second, perhaps unsurprisingly, \name helps MSCN largely reduce the constraint violations over both datasets. For example, on DSB, the ratio of violations of consistency constraint can be reduced from $7\%$ to $2.4\%$ after training with \name. Indeed, for all constraints evaluated for both datasets, we observe more than $40\%$ and on average $50 \%$ reduction in constraint violations.
\emph{The improvement of Constraint-MSCN over MSCN indicates the effectiveness of \name in reducing constraint violations for learned databases, forcing them to "acquire" the domain knowledge represented by the constraints.}

\subsection{Can \name Improve the Cardinality Estimation Accuracy?}\label{section.exp.card}

The previous section established that \name\ helps MSCN learn the effects of integrity constraints.  We now focus on the bigger picture, which is whether this translates into improved cardinality estimation accuracy --- the primary task of the learned optimizer.

\begin{table}[ht]
\centering
\scalebox{0.95}{
\begin{tabular}{|c|c|c|c|c|}
  \hline
  \multirow{2}{*}{Model}  &\multicolumn{2}{c|}{In-Dis Queries}  & \multicolumn{2}{c|}{OOD Queries}  \\
\cline{2-5}
  & Median & 95th &   Median & 95th   \\
    \hline
  {PostgreSQL} & 7.2 & 9437 & 5.0 &  192 \\
  \hline
{MSCN}  &  1.5 & 24 & 1.4 & 11   \\
  \hline
   \makecell{Constrained MSCN}  & 1.4 & 24  & 1.3  & 6.5  \\ 
   \hline
   \makecell{Constrained MSCN-50\%}  & 1.5 & 47  & 1.4  & 7.4  \\ 
  \hline
\end{tabular}}
\caption{\mbox{Estimation performance (Q-error) on IMDb}}
\label{exp.table.qerror.imdb}
\end{table}
\begin{table}[ht]
\centering
\scalebox{0.95}{
\begin{tabular}{|c|c|c|c|c|}
  \hline
  \multirow{2}{*}{Model}  &\multicolumn{2}{c|}{In-Dis Queries}  & \multicolumn{2}{c|}{OOD Queries}  \\
\cline{2-5}
  & Median & 95th &   Median & 95th   \\
    \hline
  {PostgreSQL} & 1.3 & 29 & 2.0  & 21  \\
  \hline
{MSCN}  &  1.2 & 5.1 &   1.2 & 133   \\
   \hline
   \makecell{Constrained MSCN}  & 1.2 & 4.7 &   1.2 & 7.0  \\ 
   \hline
   \makecell{Constrained MSCN-50\%}  & 1.2 &  5.5  &   1.2 &  11   \\ 
  \hline
\end{tabular}}
\caption{\mbox{Estimation performance (Q-error) on DSB}}
\label{exp.table.qerror.dsb}
\end{table}

\vspace{-1em}

Tables~\ref{exp.table.qerror.imdb} and~\ref{exp.table.qerror.dsb} present the estimation performance on IMDb and DSB, respectively.
We refer to MSCN trained with \name as Constrained-MSCN. 
From the tables, we can see that baseline MSCN outperforms PostgreSQL in most cases. However, and importantly: MSCN's tail performance may suffer on OOD queries. For example, on DSB, MSCN's q-error is more than $6\times$ (133 \textit{vs.} 5.1) worse than PostgreSQL at the 95-percentile (\textbf{95th} column). This shows the limited generalization ability of MSCN. However, \name makes Constrained MSCN much more robust to OOD queries. For example, it improves MSCN's performance by $95\%$ (to 7.0) at the same percentile on DSB. Additionally, we observe that the improvement of Constrained-MSCN over MSCN is significantly larger on OOD queries compared to that on In-Dis queries. This further justifies our argument that \name offers better generalization ability to query-driven learned models.

\subsection{Can \name Reduce the Training Set Size?} 

\begin{figure*}[ht!]
\centering
\begin{subfigure}{.235\textwidth}
\captionsetup{justification=centering}
\includegraphics[width=\textwidth]{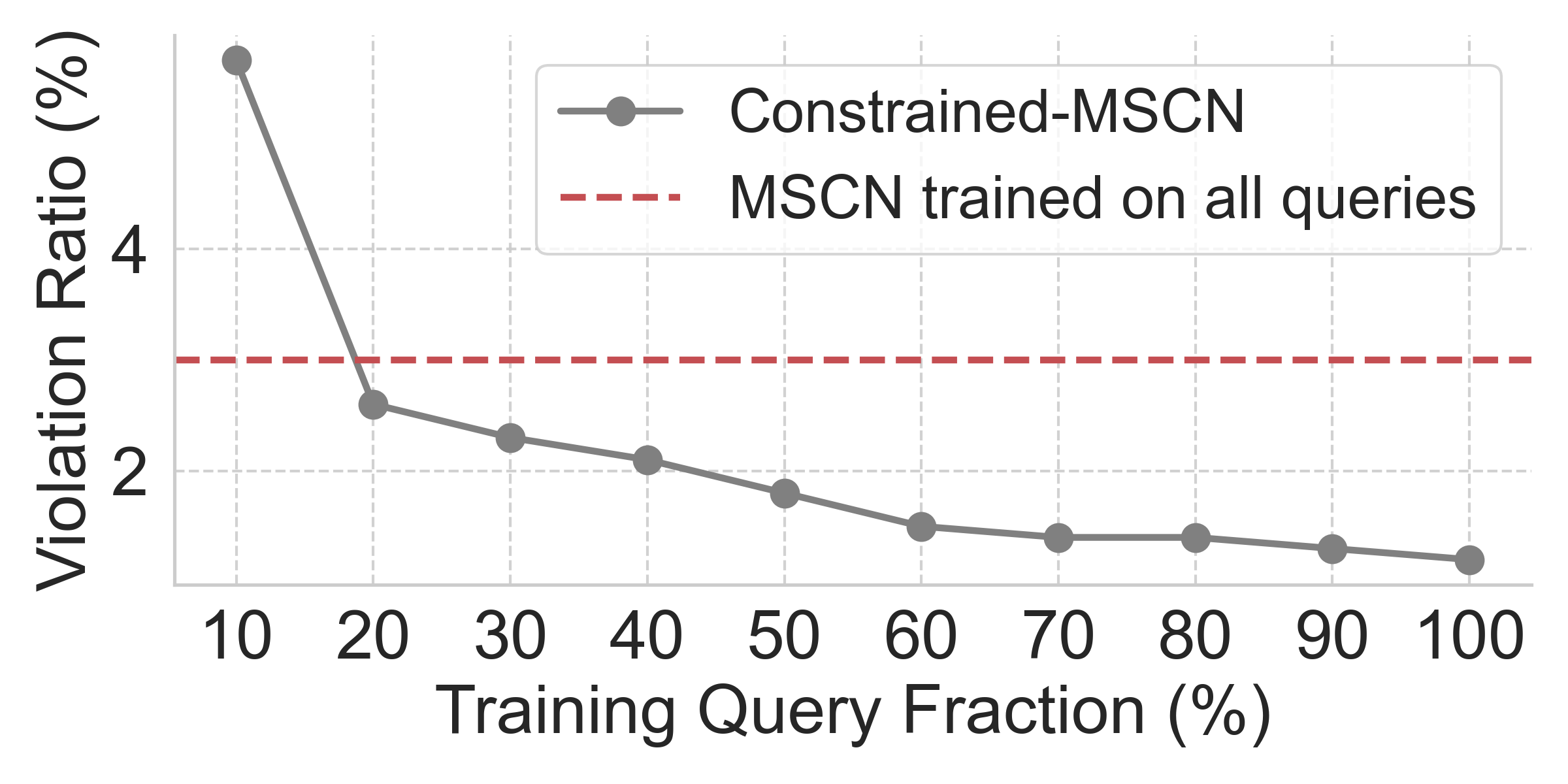}
\caption{IMDb Consistency}
\end{subfigure}
\begin{subfigure}{.235\textwidth}
\captionsetup{justification=centering}
\includegraphics[width=\textwidth]{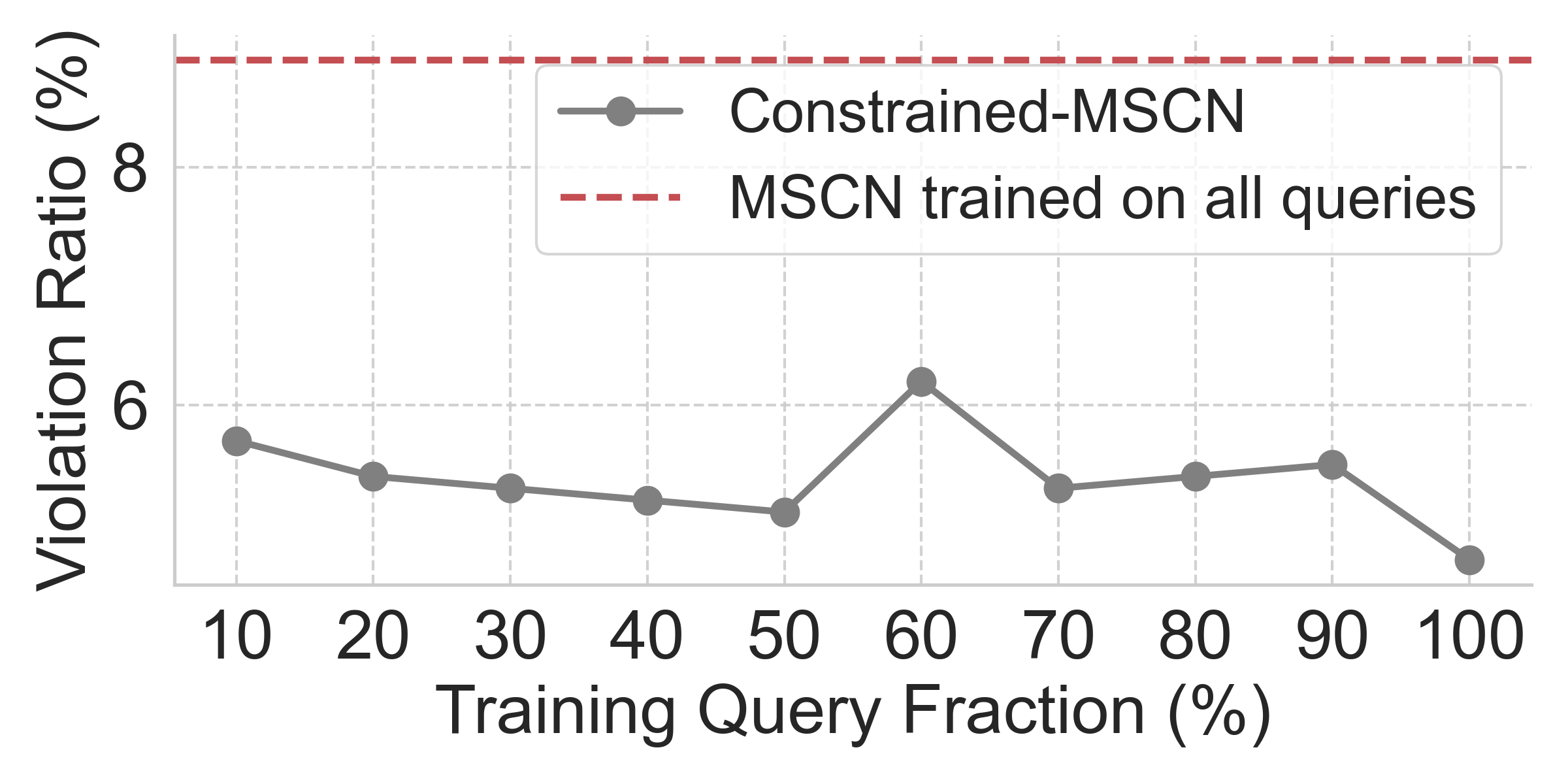}
\caption{IMDb PK-FK Inequality}
\end{subfigure}
\begin{subfigure}{.235\textwidth}
\captionsetup{justification=centering}
\includegraphics[width=\textwidth]{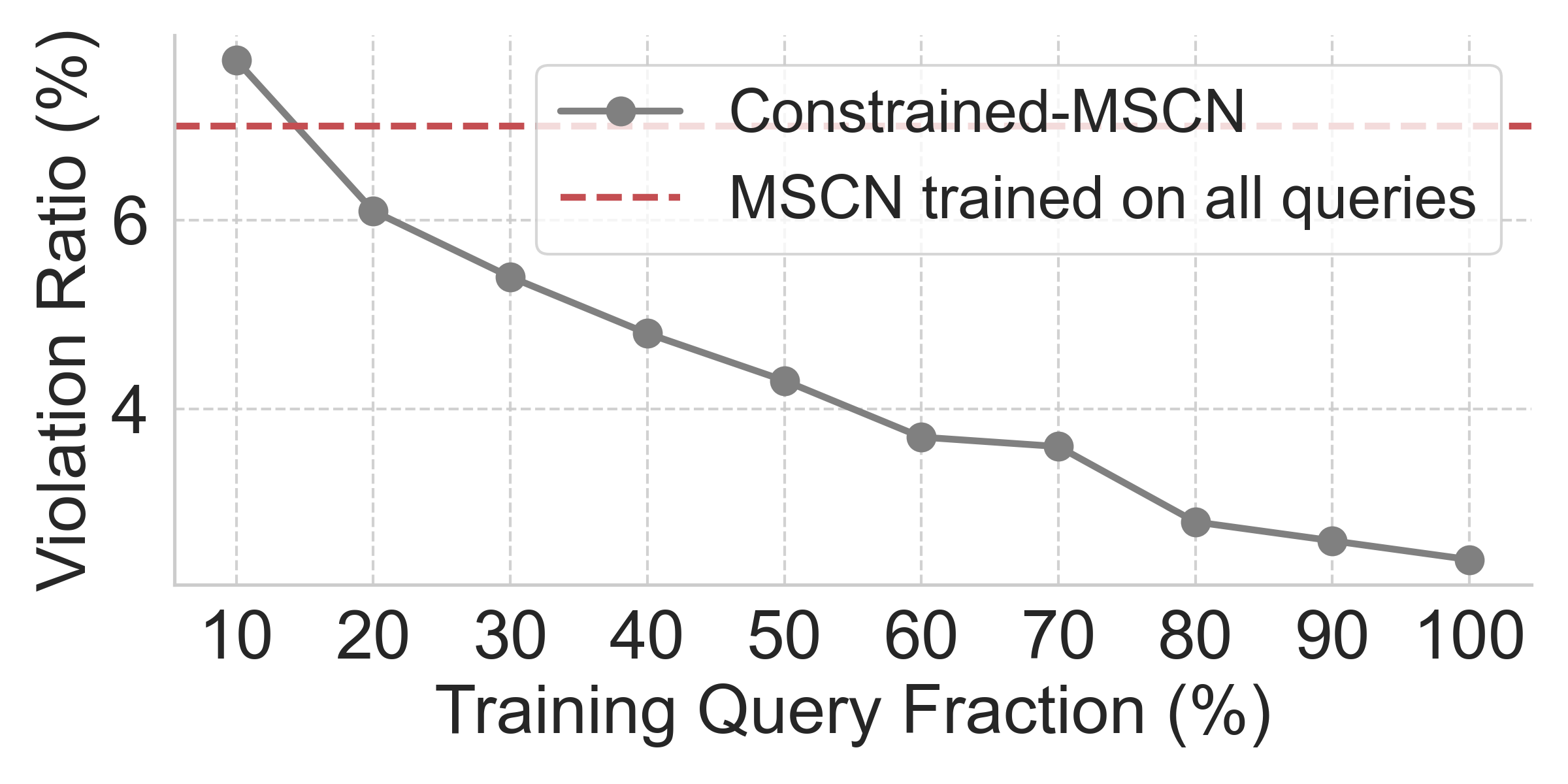}
\caption{DSB Consistency}
\end{subfigure}
\begin{subfigure}{.235\textwidth}
\captionsetup{justification=centering}
\includegraphics[width=\textwidth]{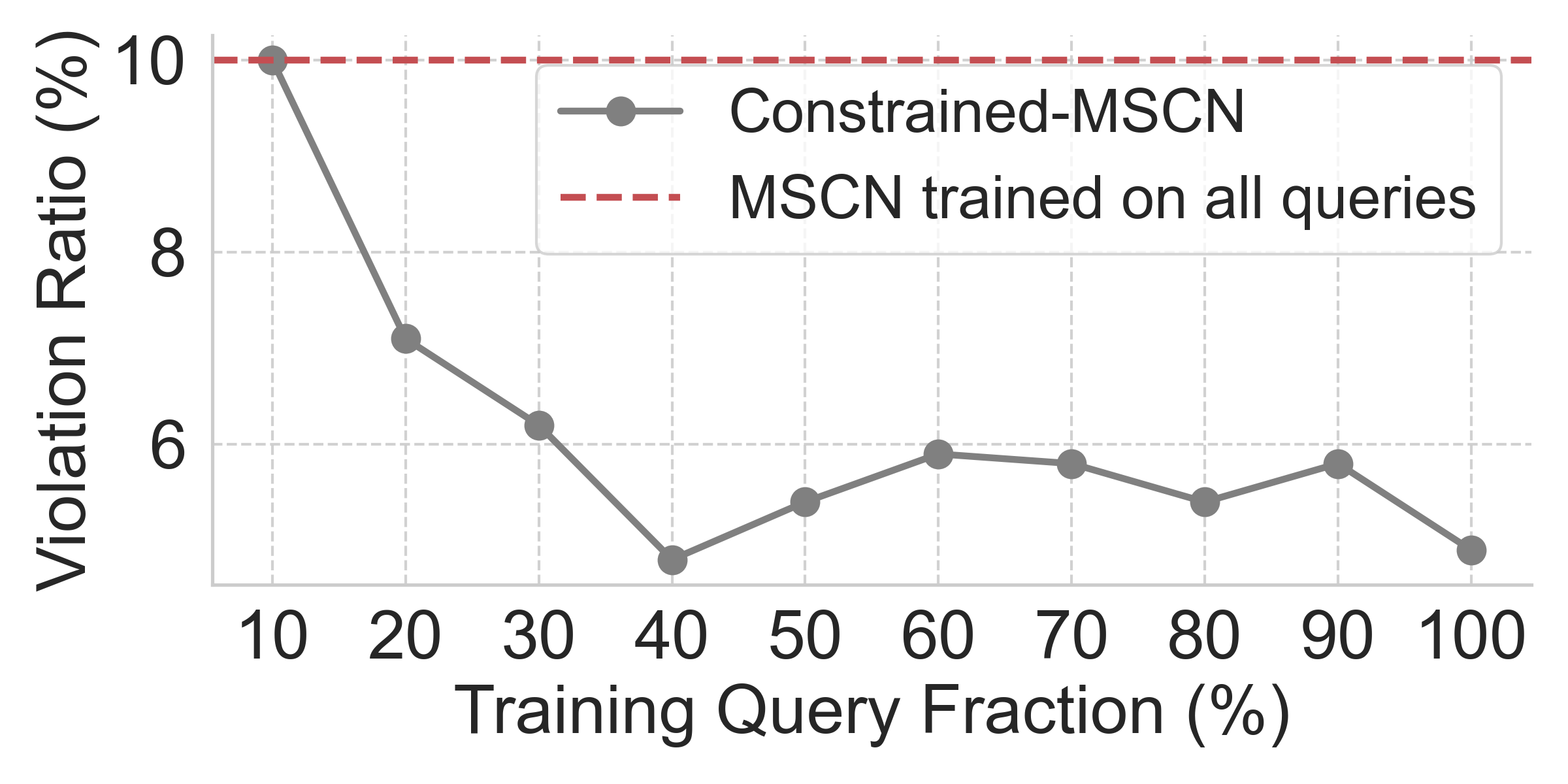}
\caption{DSB PK-FK Inequality}
\end{subfigure}
\caption{Ratios of constraint violations with varying fractions of labeled training queries for Constrained-MSCN.}\label{fig.exp.vary.vio}
\end{figure*}

\begin{figure*}[ht!]
\centering
\begin{subfigure}{.235\textwidth}
\captionsetup{justification=centering}
\includegraphics[width=\textwidth]{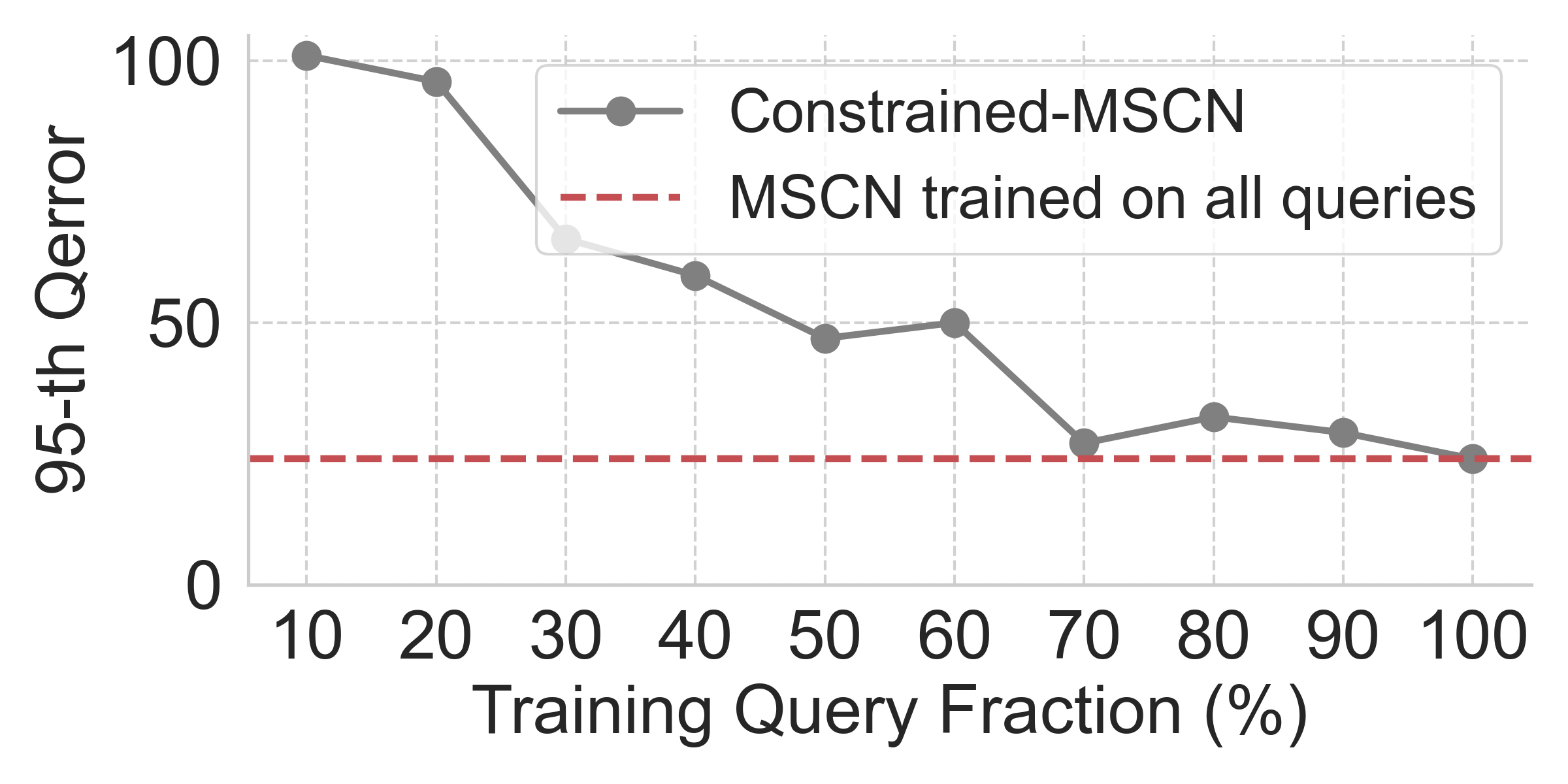}
\caption{IMDb In-Dis 95-th Qerror}
\end{subfigure}
\begin{subfigure}{.235\textwidth}
\captionsetup{justification=centering}
\includegraphics[width=\textwidth]{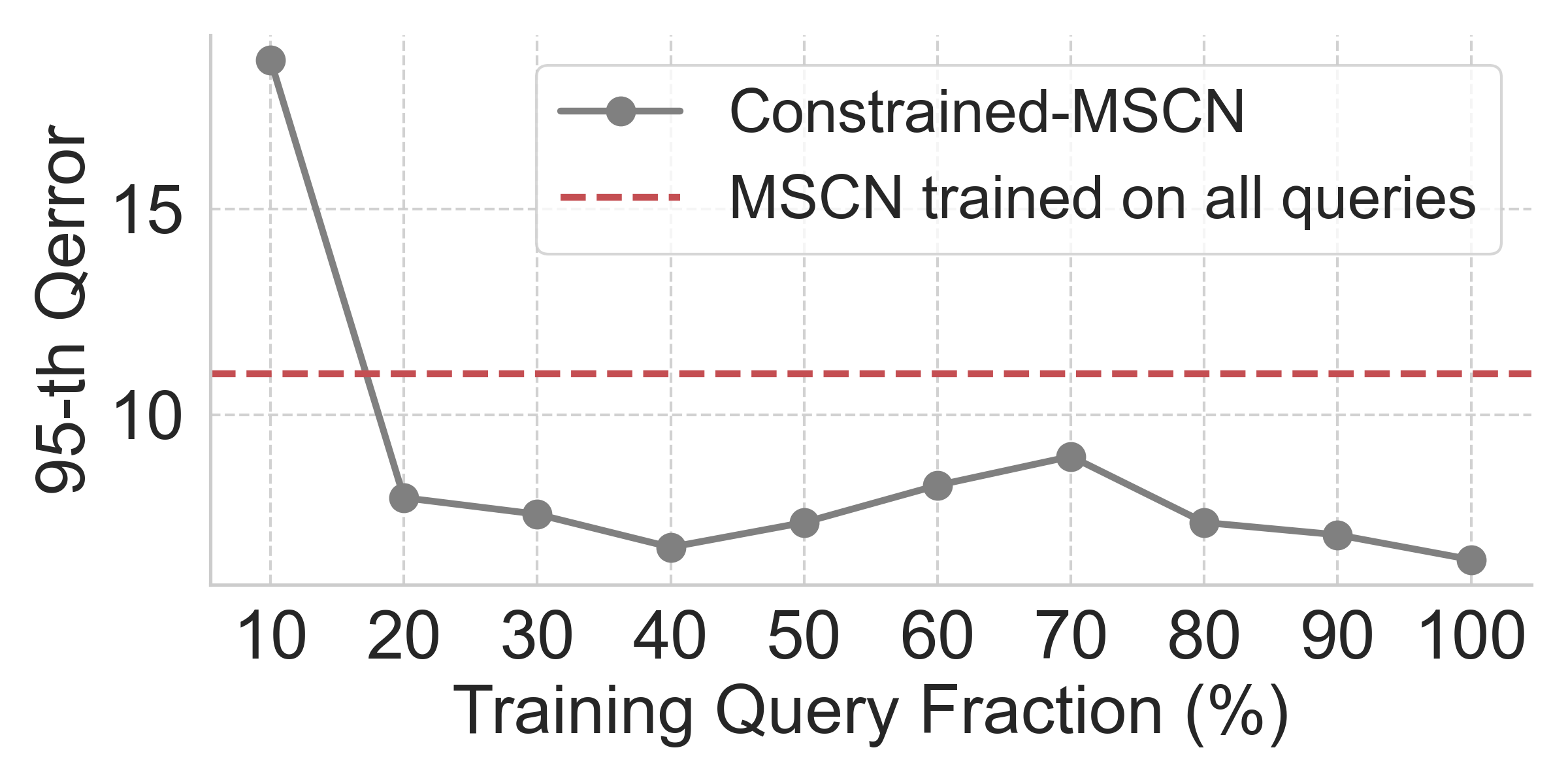}
\caption{IMDb OOD 95-th Qerror}
\end{subfigure}
\begin{subfigure}{.235\textwidth}
\captionsetup{justification=centering}
\includegraphics[width=\textwidth]{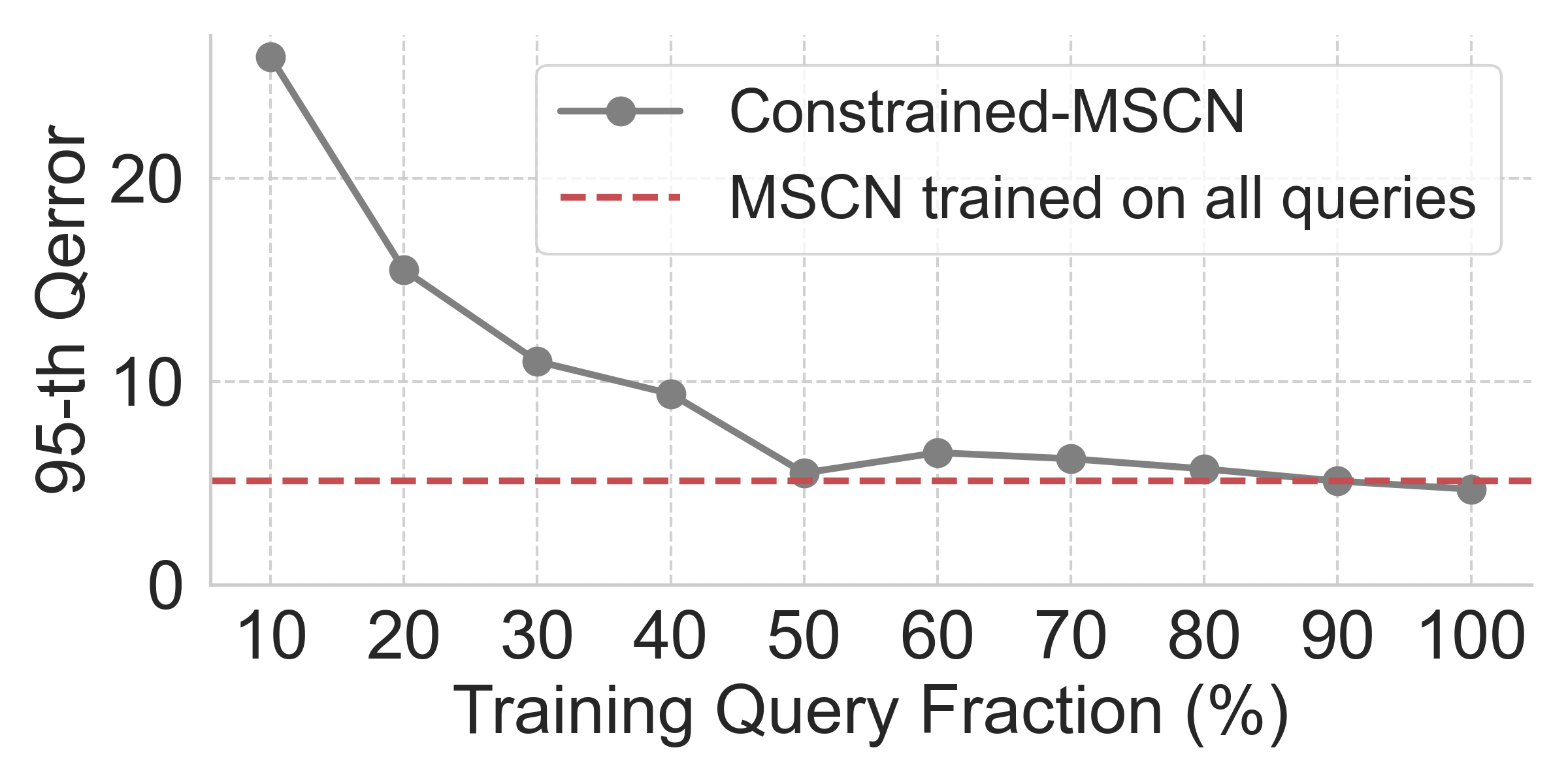}
\caption{DSB In-Dis 95-th Qerror}
\end{subfigure}
\begin{subfigure}{.235\textwidth}
\captionsetup{justification=centering}
\includegraphics[width=\textwidth]{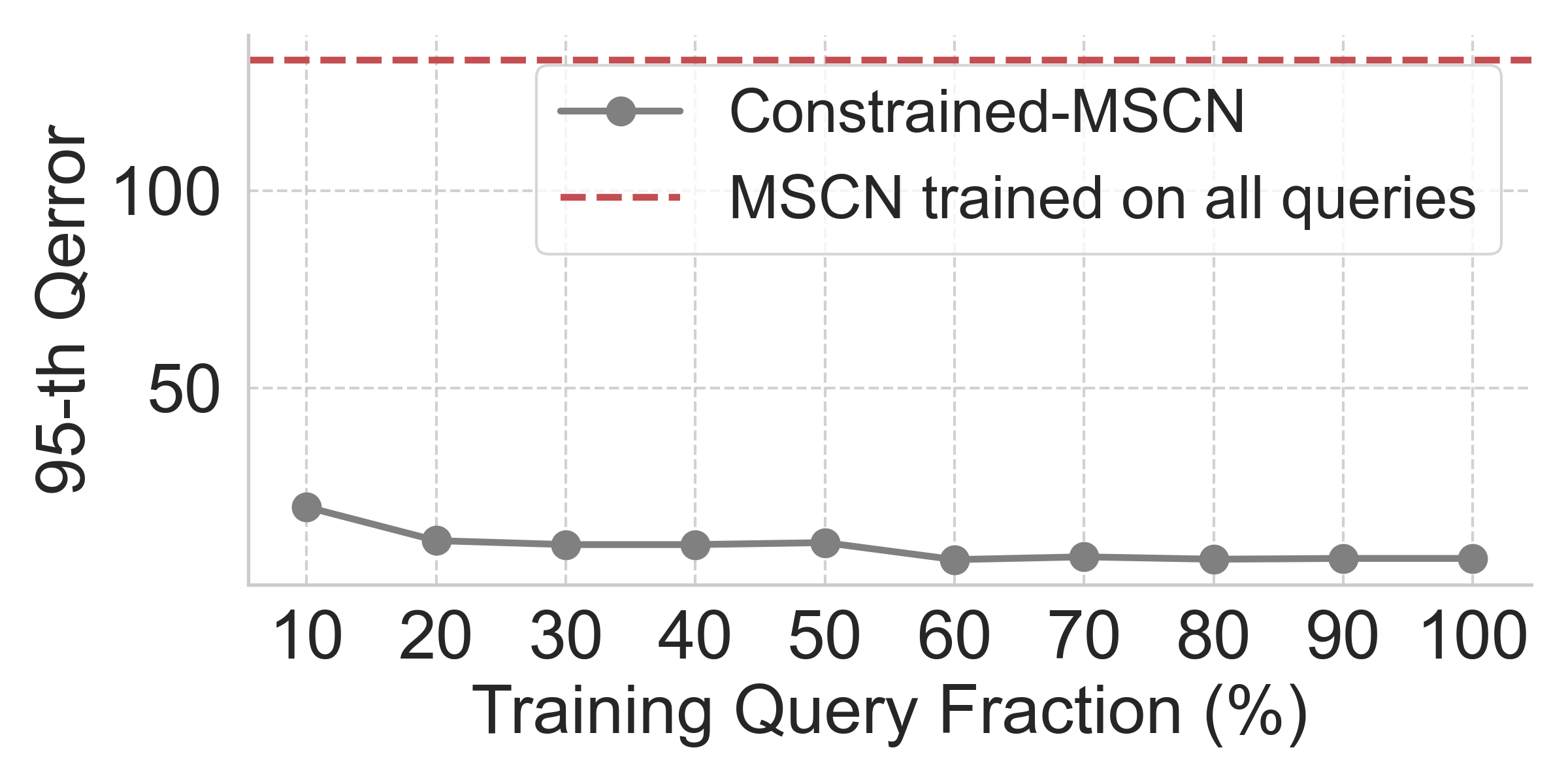}
\caption{DSB OOD 95-th Qerror}
\end{subfigure}
\caption{Cardinality estimation performance (q-error) with varying fractions of labeled training queries for Constrained-MSCN.}\label{fig.exp.vary.card}
\end{figure*}

Intuitively, if MSCN leverages domain knowledge into its cardinality estimation via \name, we might be able to use fewer training examples (which are costly to obtain) to train the model while attaining good performance.
To this end, we repeat the experiments of constrained-MSCN in Sections~\ref{section.exp.vio} and~\ref{section.exp.card}, with varying subsets of the original training set (comprised of queries and their cardinalities). Figures~\ref{fig.exp.vary.vio} and~\ref{fig.exp.vary.card} show constraint violations and cardinality estimates, respectively. Due to space constraints, we visualize the number of consistency constraint and PK-FK inequality constraint violations in Figure~\ref{fig.exp.vary.vio}, and the cardinality estimation q-errors at the 95-th percentile in Figure~\ref{fig.exp.vary.card}. For easy comparison, we also augment Tables~\ref{exp.table.numvio.dsb} and~\ref{exp.table.qerror.imdb} with more detailed results of constrained-MSCN trained on 50\% queries (which we refer to Constrained-MSCN-\%50).

With respect to constraint violations (Figure~\ref{fig.exp.vary.vio}), Constrained-MSCN better satisfies consistency constraints, compared to MSCN trained on all queries (\textit{aka.,} MSCN-Full), with only $20\%$ labeled training queries. For the PK-FK inequality constraint, Constrained-MSCN outperforms MSCN-Full from the beginning (\textbf{10\%}  labeled training queries). Additionally, from Table~\ref{exp.table.numvio.imdb} and Table~\ref{exp.table.numvio.dsb}, we can see Constraint-MSCN-50\% has consistently smaller constraint violations across \emph{all} constraints on both datasets. Clearly, it learns to respect these constraints with only a small subset of the training queries.

Cardinality estimation (Figure~\ref{fig.exp.vary.card}) also achieves good accuracy with notably less data. For In-Dis queries, Constrained-MSCN can achieve the same performance as MSCN-Full using 50\% to 70\% labeled training queries. Perhaps interestingly, and more importantly, we observe Constrained-MSCN can outperform MSCN-Full with only \textbf{20\%} labeled training queries for both datasets' OOD queries. Notably, on the DSB dataset, Constrained-MSCN is $5$ times better than MSCN-Full even trained on only 10\% labeled training queries.

The empirical results demonstrate that \name can help query-driven learned cardinality models achieve better performance, especially in generalization ability, \emph{with a significantly reduced number of labeled training queries.} This underscores the \emph{practicality} of \name since labeled queries can be very expensive to obtain.

\subsection{Does \name Translate into a Better End-to-End Performance?}

In this subsection, we demonstrate that incorporating domain knowledge with \name can lead not only to better cardinality estimates, but also to a better end-to-end (running time) performance. All end-to-end experiments are conducted with a modified version of PostgreSQL 13.1 that can accept injected cardinalities estimates~\cite{pgcode,ceb}. We compare constrained-MSCN to the original, PostgreSQL (histogram-based estimates, a baseline upon which learned cardinality estimation should improve) and True cardinalities (the best performance to which the learned models could be optimized). We describe our results in detail below.

\subsubsection{\textbf{Performance on Benchmark Queries}} \label{section.exp.run.bechmark}
We first evaluate the performance with a standard benchmark JOB-light for the IMDb dataset. Interestingly, we observe significant variations in the performance of MSCN on the JOB-light benchmark between two recent benchmarking papers~\cite{han2021CEbenchmark,kim2022learned}: MSCN performance is slightly better than PostgreSQL in~\cite{han2021CEbenchmark}, whereas it is significantly better than PostgreSQL (and close to true cardinalities) in~\cite{kim2022learned}. We reproduced their experiments and confirmed the results: the gap is due to the different properties of training workloads, as we discuss in Section~\ref{section.exp.setup}. This also justifies our choice of training workload for IMDb. For DSB, we sample 50 queries with more than 2 joins from the test queries to construct the DSB-50 workload.

\begin{figure}
\centering
\begin{subfigure}{.235\textwidth}
\captionsetup{justification=centering}
\includegraphics[width=\textwidth]{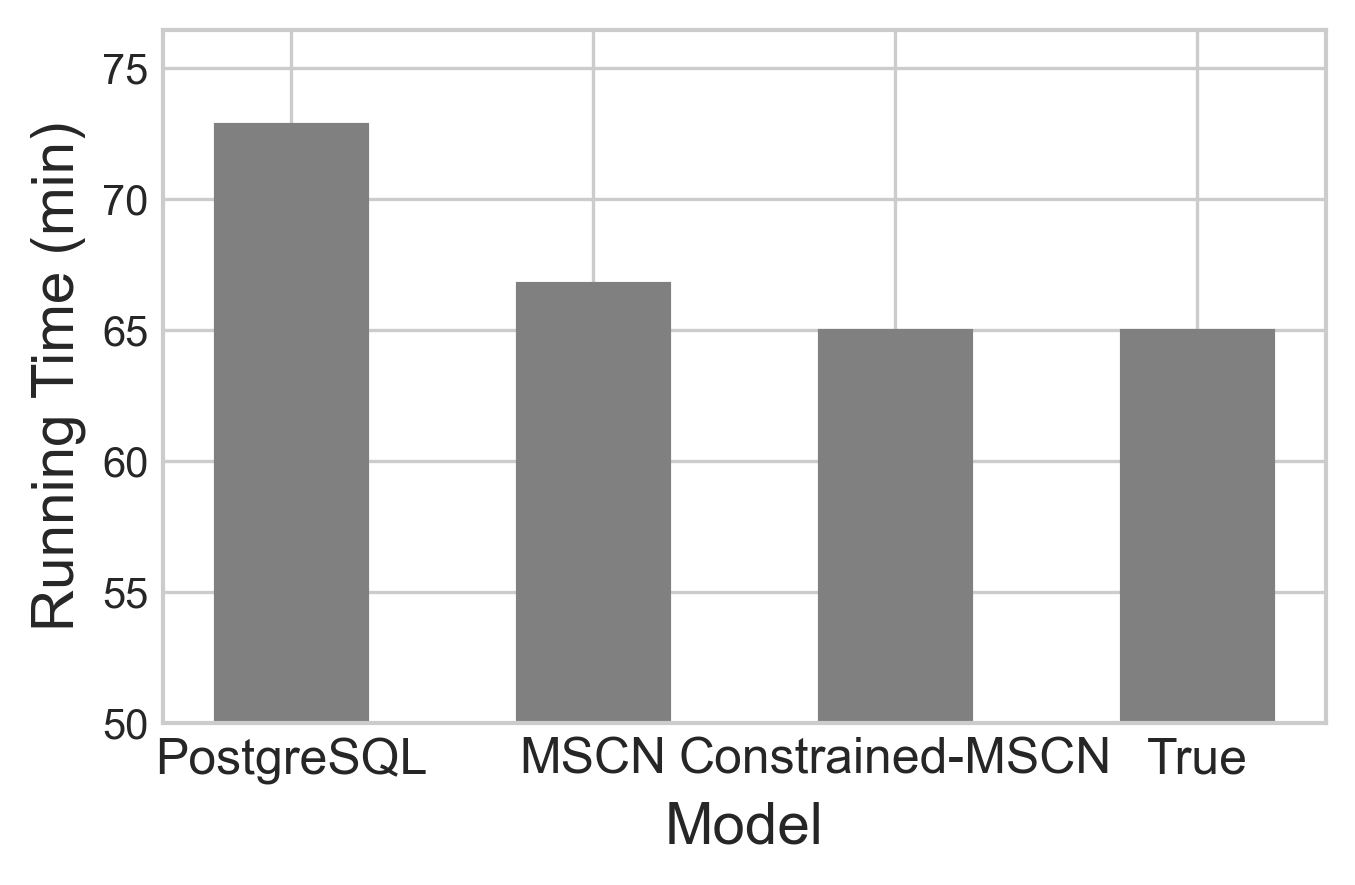}
\caption{JOB-Light Workload}
\end{subfigure}
\begin{subfigure}{.235\textwidth}
\captionsetup{justification=centering}
\includegraphics[width=\textwidth]{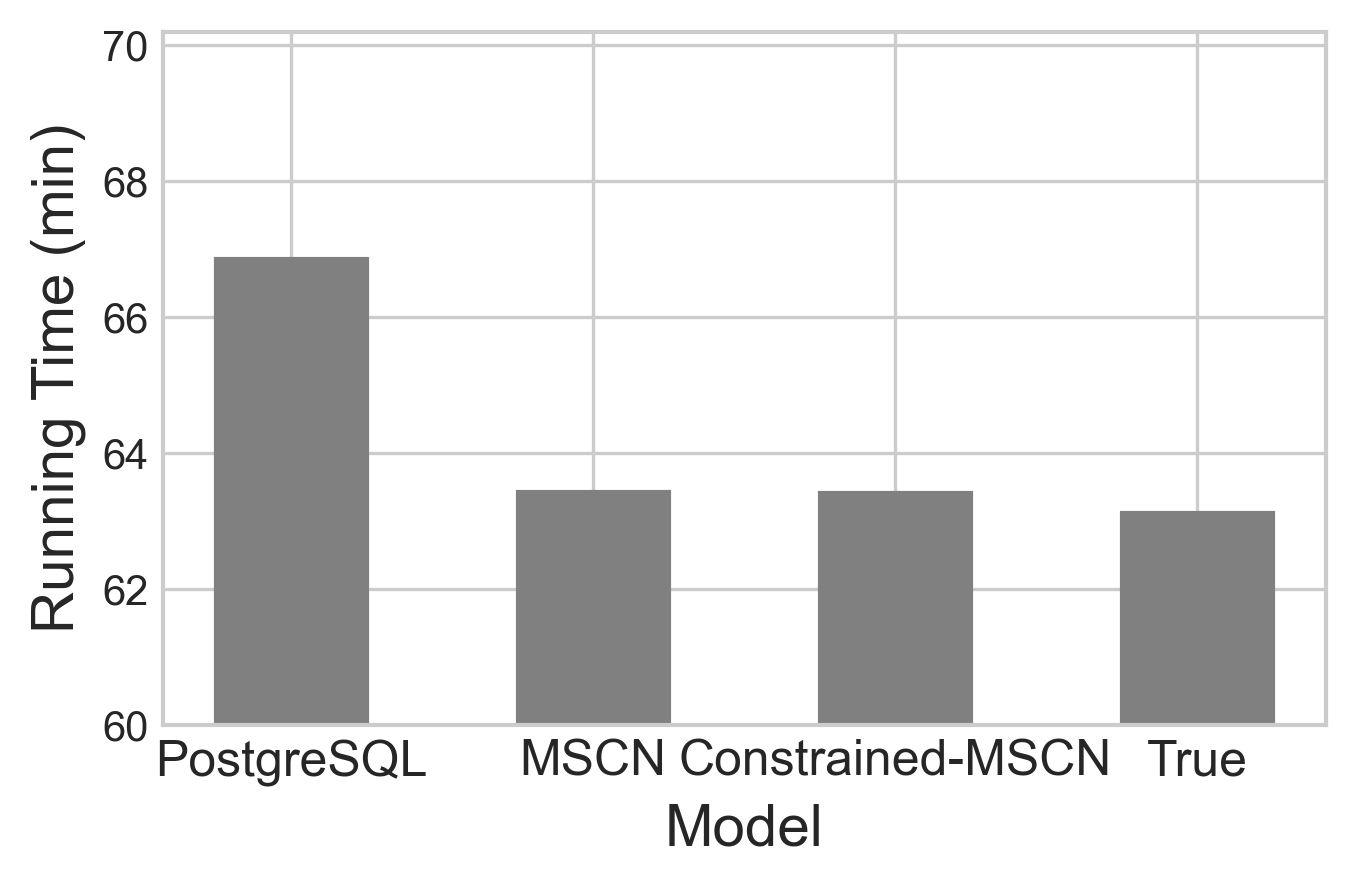}
\caption{DSB-50 Workload}
\end{subfigure}
\caption{End-to-end performance on benchmark queries.}\label{fig.exp.time.normal}
\end{figure}

Figure~\ref{fig.exp.time.normal} presents the end-to-end performance on benchmark queries. From the comparison of PostgreSQL and query-driven learned models (\textit{e.g.,} MSCN and constrained-MSCN), we can see that \emph{MSCN, if trained on a proper workload, could consistently outperform PostgreSQL on benchmark queries}. In other words, MSCN and constrained-MSCN can achieve a significant running time reduction compared to PostgreSQL. Indeed, they are even on par with the performance of true cardinalities.  However, end-to-end benefits on overall workload running times are largely masked by the fact that the benchmarks contain a mix of queries (not all of which have key or range constraints), and MSCN exhibits strong overall performance on these.

\subsubsection{\textbf{Performance on Queries Requiring Domain Knowledge}} 
To further measure the impact of \name's ability to add domain knowledge to query-driven models, we propose to focus not on benchmark queries, but rather on the "domain knowledge-sensitive" queries (which we refer to DKS queries) where standard query-driven learned query optimization does poorly due to its lack of domain knowledge. 


Naturally, given a query workload that consists of a set of $n$ \emph{candidate queries}, we can retrospectively analyze the query workload to determine where a learning-based optimizer has performed poorly. However, this requires a trained model and executing all $n$ \emph{candidate queries} with the model's estimates. We propose that, as a building block towards evaluation of our work, as well as future work that exploits domain knowledge --- and potentially for selecting candidates for improved training --- it is useful to have an algorithmic approach to efficiently identifying
DKS queries \emph{a priori}, \emph{based on domain knowledge and the violations of constraints}, without actually executing all $n$ queries. 
Our method is simple, principled, and general which can be applicable to different kinds of constraints.  We first define the goal of the algorithmic approach.

\medskip

\noindent \textbf{Goal.}
Given a set of $n$ \emph{candidate queries}, we seek to find the top-$k$ \emph{DKS queries} for which the learned query optimizers are \emph{most likely} to produce bad query execution plans which subsequently lead to slower query running time. For our paper's focus on learned cardinality estimation, our task is to find the \emph{DKS queries} for which the learned cardinality model will most likely give \emph{bad estimates} that result in a bad query running time performance.  We develop this into a \emph{new benchmark for learned query optimization}.

\smallskip

\noindent \textbf{Approach.} \label{section.attack.steps}
 The key idea is \emph{identifying queries with largely underestimated subqueries by the learned cardinality model as the DKS queries}. The rationale for this idea is intuitive and justified by previous work~\cite{leis2018query, deeds2023safebound,han2021CEbenchmark}: underestimation leads optimizers to produce overly optimistic query plans whose costs are very high for expensive
queries. The proposed approach has three steps. First, we split each query into a set of all subqueries. Second, we measure the degrees of underestimation for all subqueries by the learned cardinality model using constraints. Third, we rank original queries based on the value of the \emph{largest} underestimation degree among their subqueries and return the top $k$ original queries as the DKS queries. Here the key issue is how to estimate the degrees of model underestimation via constraints. Next, we will discuss the cases for the two constraints (\textit{e.g.,} consistency constraint and PK-FK equality constraint) that we apply in the construction of DKS queries in this paper. Note that the entire approach does not require any query execution, which could be very expensive.

\smallskip
\noindent \textbf{Finding DKS Queries via Consistency Constraints.} \label{section.find-adv}
Our first subtask is to use consistency constraints to estimate the model underestimation degrees for all subqueries.
For each subquery $q_i$, We use the approach discussed in Section~\ref{sec:con_consistency} to break $q_i$ into two separated queries $q_{i1}$ and $q_{i2}$.

Afterward, we use the learned cardinality model to estimate the cardinalities for the three queries ($q_i$, $q_{i1}$, and $q_{i2}$).  We measure the model underestimation degrees based on two insights. First, if $|\hat{q_i}| < |\hat q_{i1}| + |\hat q_{i12}|$, we know it is a clear violation of consistency constraint for the learned cardinality model. Second, the higher the ratio $\frac {|\hat q_{i1}| + |\hat q_{i12}|} {|\hat q_i|}$ is, the higher the likelihood of $q_i$ being underestimated by the model, so we call it \emph{underestimation ratio}. 
As such, we use the underestimation ratio as the predicted underestimation degree to sort all subqueries. Here the intuition is that for a predicted underestimation degree that is sufficiently large, there is a high chance of $q_{i}$ being underestimated even if $q_{i1}$ and $q_{i2}$ are overestimated.



\smallskip
\noindent \textbf{Finding DKS Queries via PK-FK Equality Constraints.}  We follow Section~\ref{sec:con_joins} to generate query $q_B$ for each query $q$, $|q| = |q_B|$ if there is no predicate on $q_B$. 
Then, similar to the consistency constraint, we get their model predictions, $|\hat q|$ and $|\hat q_B|$ and use the ratio $|\hat q_B| / |\hat q|$ as the predicted underestimation degree. The intuition is likewise to that of consistency constraint --- the higher this ratio is, the more likely $q$ is underestimated.

Astute readers may be concerned that the PK-FK equality constraints are seldom useful in practice. In fact, the applicable scenario does not compromise the practicality of the PK-FK equality constraint. First, join queries are much more important than single table queries in query optimization. Second, in fact, in the existing query optimization benchmarks~\cite{han2021CEbenchmark, leis2015good} it is common to see queries with no predicate on the fact table. Third, for IMDb dataset, even for those queries having predicates on the fact table, a large fraction of their subqueries contains only FK-FK join. For these subqueries with only FK-FK join, MSCN (and many other learned cardinality methods such as DeepDB~\cite{deepdb}) need to first add the fact table (with no predicate) to them and convert the FK-FK join to multiple PK-FK joins (which will not change the query results), and then estimate the resulting subqueries. The new subqueries precisely meet the applicable criteria of using the PK-FK constraint.




\begin{figure*}[ht!]
\centering
  \begin{subfigure}{1\textwidth}
  \centering
  \includegraphics[width=1.\linewidth]{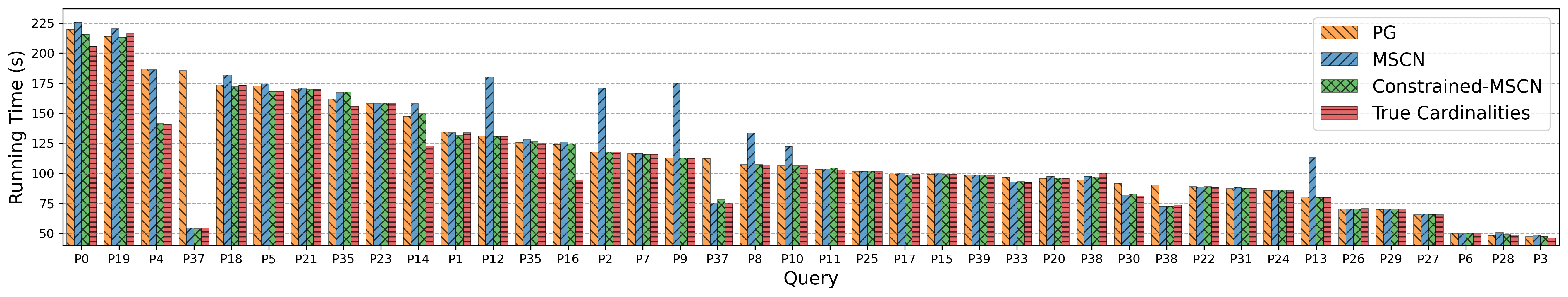}
  \caption{Chosen by PK-FK equality constraint, sorted by PostgreSQL running time} \label{fig.attack.dsb.time.fk}
\end{subfigure}

\begin{subfigure}{1\textwidth}
  \centering
  \includegraphics[width=1.\linewidth]{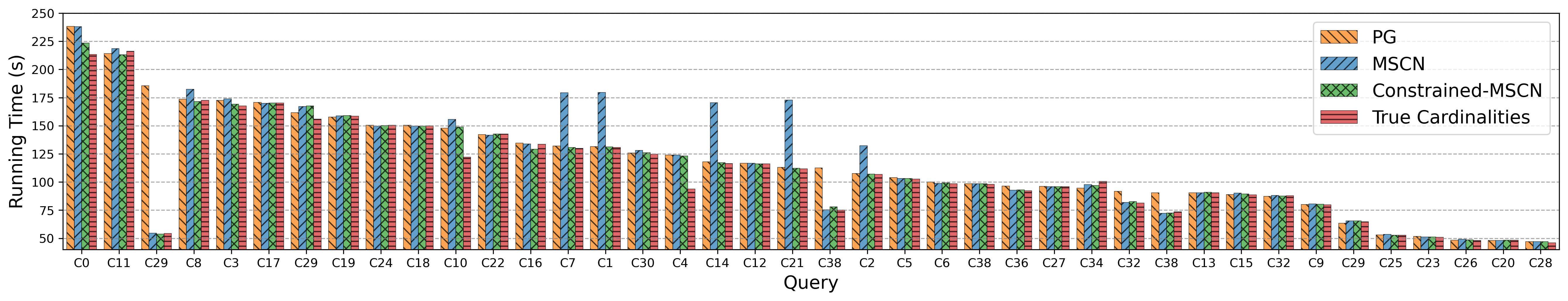}
  \caption{Chosen by consistency constraint, sorted by PostgreSQL running time} \label{fig.attack.dsb.time.cc}
\end{subfigure}%
 \caption{End-to-end (running time) performance on DSB domain knowledge-sensitive (DKS) workload.}  \label{fig.attack.time.dsb}
\end{figure*}

\begin{figure}[ht!]
\centering
  \includegraphics[width=1.\linewidth]{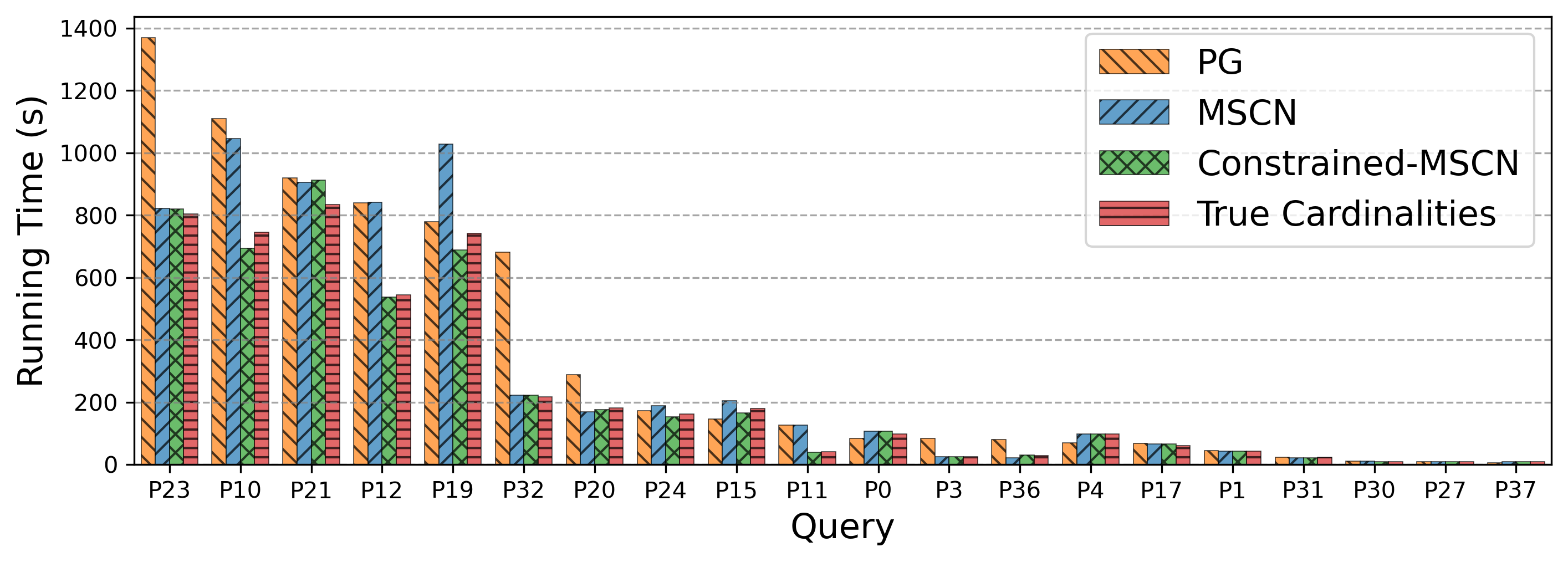}
 \caption{End-to-end performance on IMDb DKS workload chosen by PK-FK equality constraint.}  \label{fig.attack.time.imdb}
\end{figure}

\smallskip
\noindent  \textbf{Evaluation Protocol.} We first randomly sample a set of candidate queries from \emph{In-Dis} queries. Then we apply the algorithmic approach described above to identify top-$k$ DKS queries from candidate queries via the two constraints for both datasets. For DSB dataset, we mix the top 30 DKS queries returned by the approach with 10 random queries sampled from DSB-50 to form the \emph{domain knowledge-sensitive test workload} (which we refer to as the \textbf{DKS workload}). For the IMDB dataset, we directly use the top 40 DKS queries as the DKS workload.
We evaluate their query running times with PostgreSQL, true cardinalities, MSCN estimates, and Constrained-MSCN estimates, respectively. Intuitively, one may expect that PostgreSQL and true cardinalities provide the upper and lower bounds of query running time for the MSCN estimates. Note that we exclude queries with more than 4 joins (or 5 tables) from the candidate queries due to their potentially very long running time. Moreover, as we will show later, both 1) the vulnerability of query-driven MSCN due to lack of domain knowledge and 2) the advantage of Constrained-MSCN over both MSCN and PostgreSQL are \emph{already} evident on these queries considered in the evaluation, which are relatively easier to handle in query optimization compared to very-many-joins queries.

\smallskip
\noindent  \textbf{Results and Takeaways.} 
Figure~\ref{fig.attack.time.dsb} presents the query end-to-end running time performance on DSB DKS workloads chosen by PK-FK equality constraints and consistency constraints, respectively. For reference, we also include the performance of default PostgreSQL (PG) estimates. We also show the running time performance on the IMDb DKS workload chosen by PK-FK equality constraint in Figure~\ref{fig.attack.time.imdb} (we plot the first 20 queries in the order of PostgreSQL running time since the remaining 20 queries' running times are less than 1s). Note that we do not report the performance on the IMDb DKS workload chosen by consistency constraints, because we observe all compared methods are very similar (close to true cardinalities). This is likely because MSCN already performs relatively well in complying with the consistency constraint (3\% violation ratio) on the IMDb dataset (although \name can reduce the ratio by half). Nevertheless, notable enhancements of Constrained-MSCN over both PostgreSQL and MSCN are evident in the remaining three DKS workloads, underscoring the distinct value of \name.  

 Specifically, we observe three main findings from both figures. First, we observe a similar finding as in Section~\ref{section.exp.run.bechmark}, \textit{i.e.,} PostgreSQL could be improved by query-driven learned models. Second, and more interestingly, vanilla MSCN suffers from a large regression for the DKS queries compared to the default PostgreSQL on DKS workloads, for example, P12 and P2 in Figure~\ref{fig.attack.dsb.time.fk}, C7 and C1 in Figure~\ref{fig.attack.dsb.time.cc} and P19 in Figure~\ref{fig.attack.time.imdb}. Third, and perhaps not surprisingly, we observe that Constrained-MSCN combines the best of both worlds --- it can maintain the advantages of MSCN over PostgreSQL (\textit{e.g.,} P23 in Figure~\ref{fig.attack.time.imdb}) and mitigate the query regression of MSCN due to lack of domain knowledge (\textit{e.g.,} P2 in Figure~\ref{fig.attack.dsb.time.fk}). The superior results of Constrained-MSCN indicate that \emph{\name can mitigate the issue caused by the lack of domain knowledge, and translate into better end-to-end performance for MSCN}.

 \smallskip
\noindent  \textbf{Public Repository.}  To facilitate future research studies built upon \name, we are releasing the code, data, as well as the DSK benchmark used in the paper. They will be made available through the GitHub repository\footnote{\href{https://github.com/pzwupenn/CORDON/}{https://github.com/pzwupenn/CORDON/}}.

\subsection{Ablation Studies}
In this subsection, we conduct ablation studies to better understand the behaviors of the two optimization strategies in Section~\ref{section.qsampling}.

\subsubsection{\textbf{Training Methods for Multiple Constraints}} We evaluate the random choice-based method in \name for training multiple constraints in \name. We first replace the random-choice strategy in \name with \emph{applying all application constraints} (we refer to ALL) to train the Constrained-MSCN and compare the performance of Constrained-MSCN with the default random-choice strategy, in terms of both model estimation accuracy and training efficiency. The training epoch is set to 40 for all experiments. 

\begin{figure}[ht]
\centering
\begin{subfigure}{.235\textwidth}
\captionsetup{justification=centering}
\includegraphics[width=\textwidth]{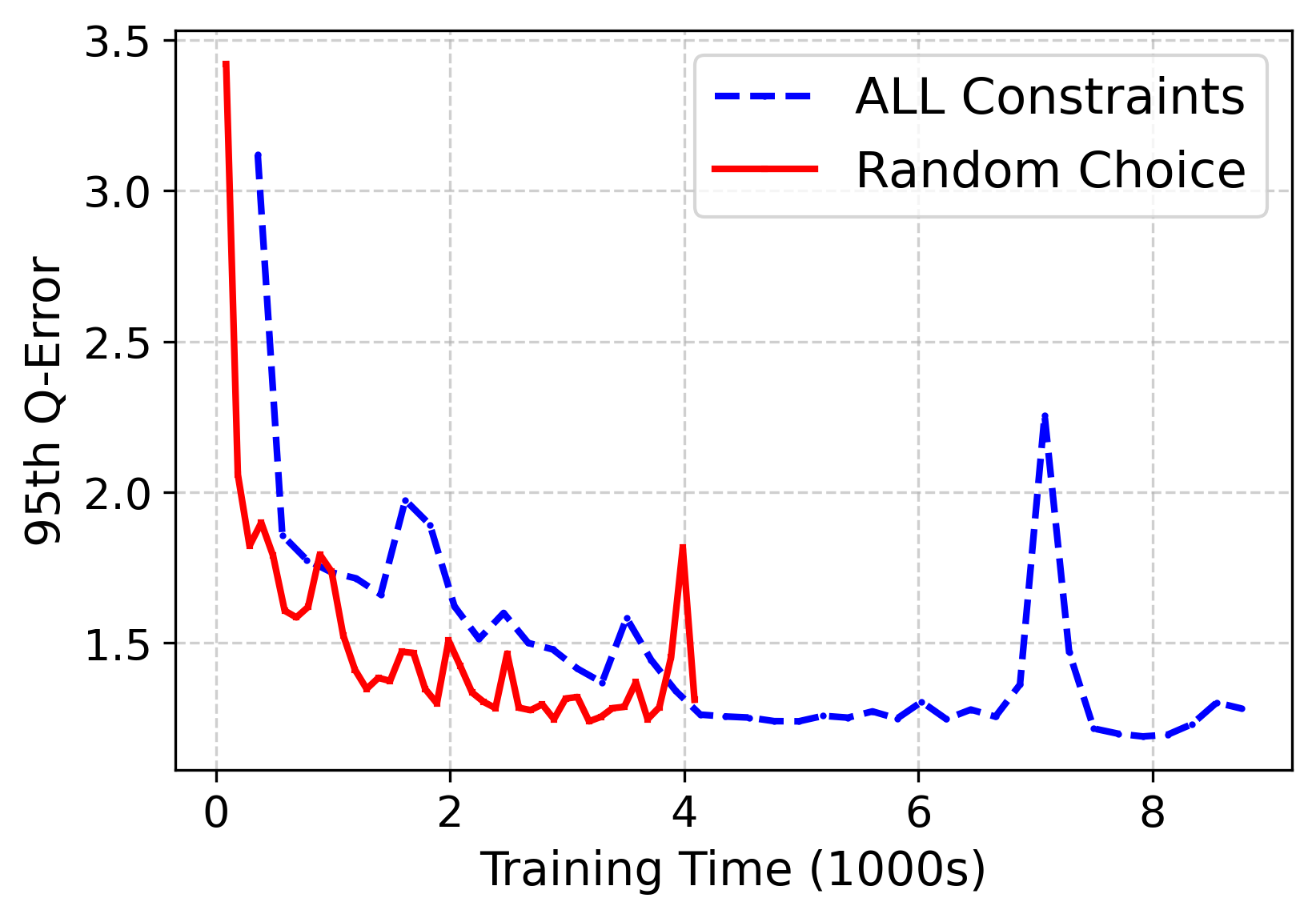}
\caption{Median: In-Dis Queries}
\end{subfigure}
\begin{subfigure}{.235\textwidth}
\captionsetup{justification=centering}
\includegraphics[width=\textwidth]{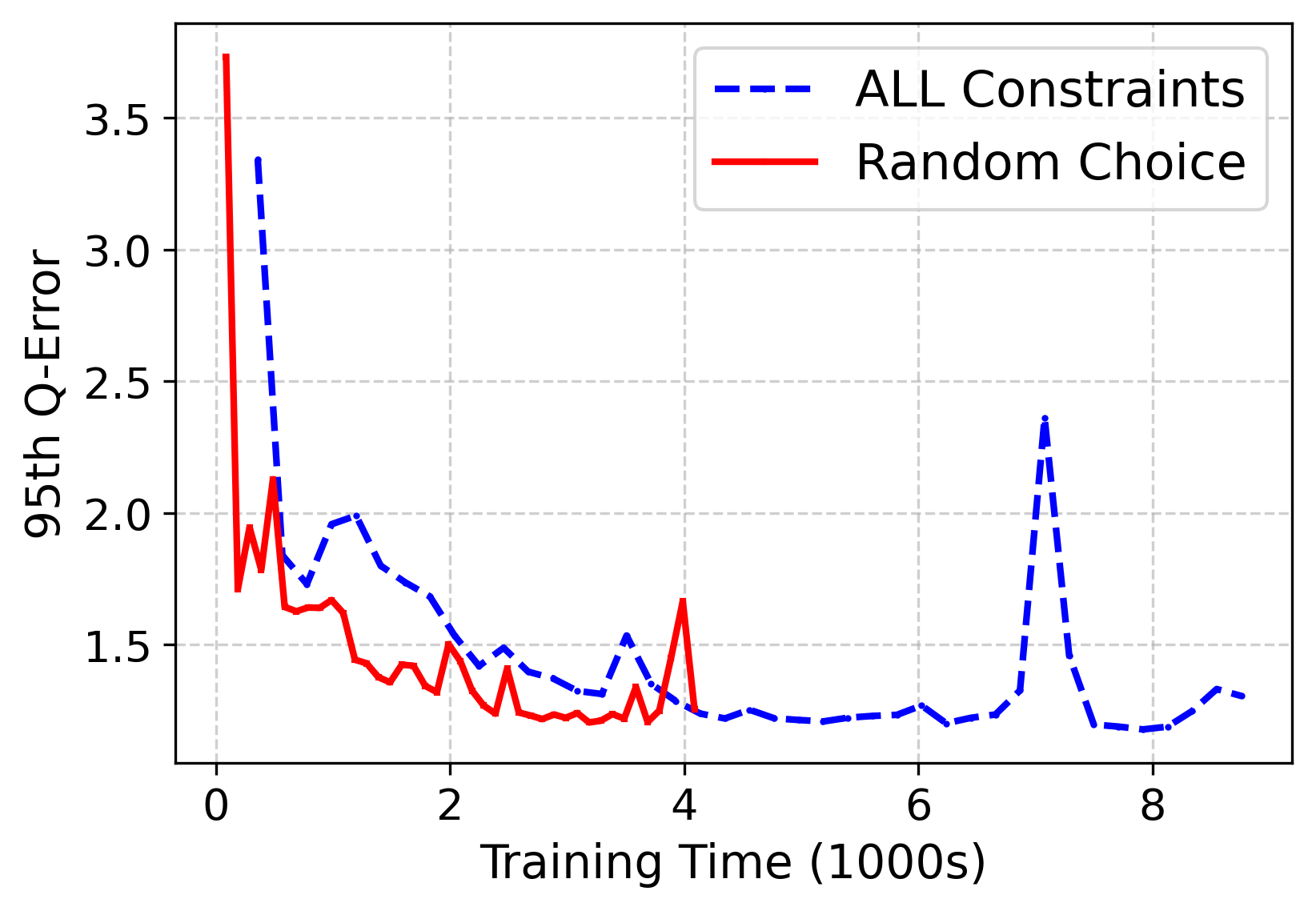}
\caption{Median: OOD Queries}
\end{subfigure}

\begin{subfigure}{.235\textwidth}
\captionsetup{justification=centering}
\includegraphics[width=\textwidth]{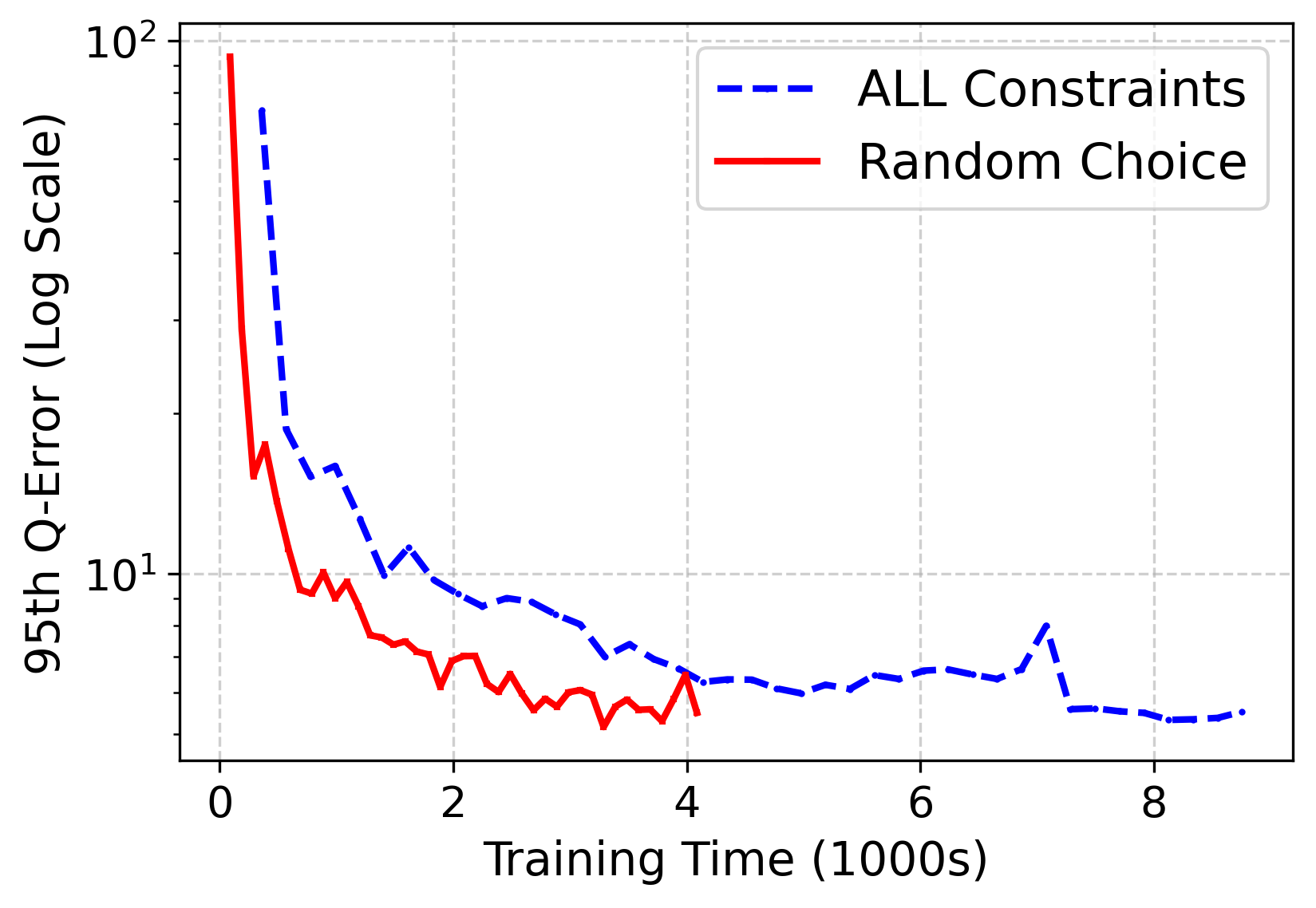}
\caption{95th: In-Dis Queries}
\end{subfigure}
\begin{subfigure}{.235\textwidth}
\captionsetup{justification=centering}
\includegraphics[width=\textwidth]{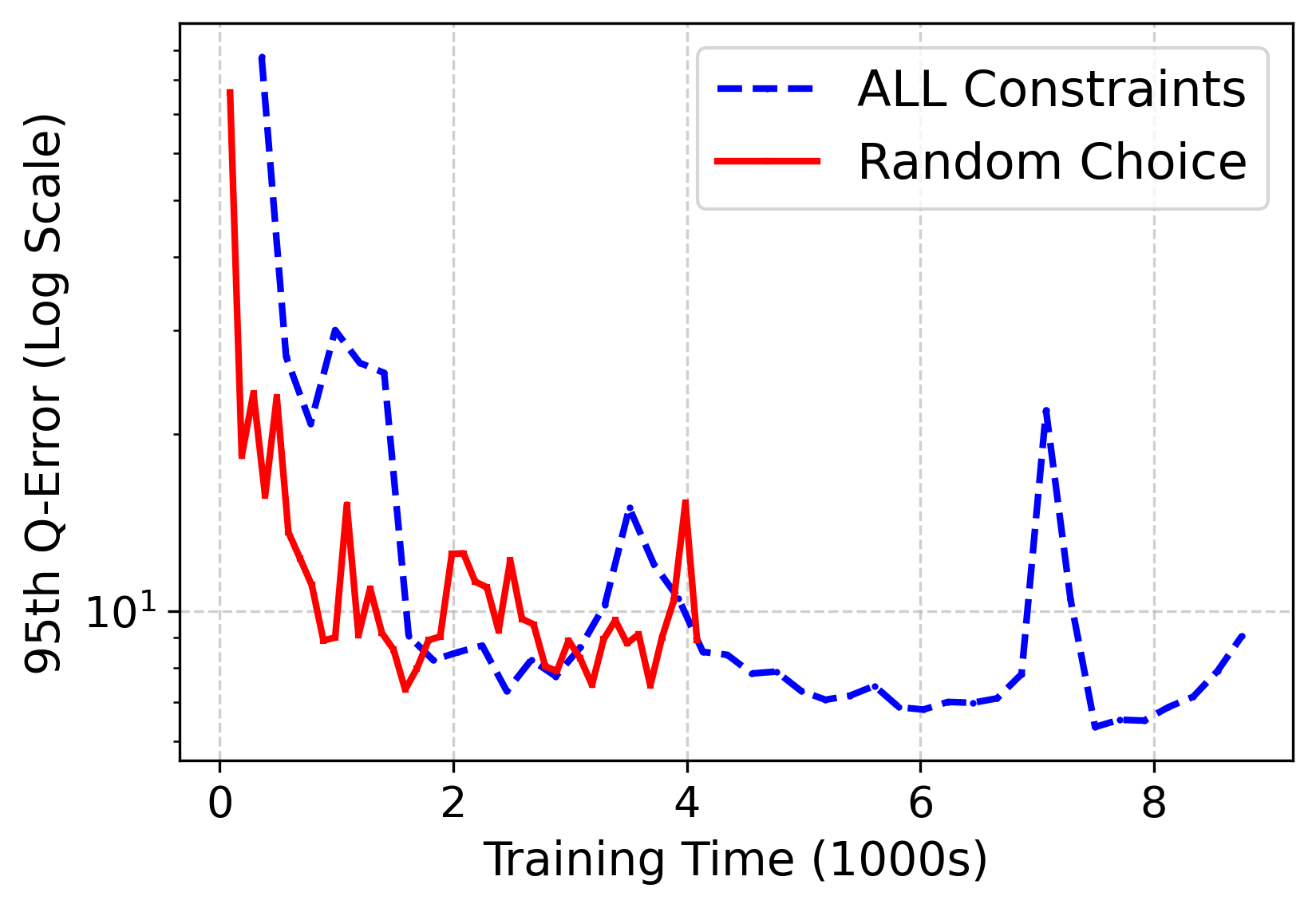}
\caption{95th: OOD Queries} \label{fig.abalation.chooseconst-d}
\end{subfigure}
\vspace{-2em}
\caption{Training curves of MSCN with \name with different methods for incorporating multiple constraints on DSB. The upper two show the median q-error and the lower two show the 95-th q-error for In-Dis and OOD queries.}\label{fig.abalation.chooseconst}
\end{figure}

Figure~\ref{fig.abalation.chooseconst} presents the learning curve of estimation accuracy (q-error) with the training time (in seconds). We show the median and 95-th percentile q-error for both In-Dis queries and OOD queries on the upper and lower levels, respectively.
From these figures, we observe that Random Choice's learning curves converge faster than those of ALL. Specifically, the average training time per epoch is 100s for random choice, less than 50\% of ALL's training time (210s per epoch). On the other hand, we observe that both training methods achieve very close accuracy at the end of training. We thus conclude that \emph{the random choice-based training strategy in \name achieves a better accuracy-efficiency trade-off than using all constraints}.


\subsubsection{\textbf{Labeling Strategies in Inequality Constraints}} Recall that we apply the pseudo-labeling mechanism to the PK-FK inequality constraint. Here we evaluate its behaviors in \name. We initiate two variations  (the default one which uses pseudo-labeling and another one without pseudo-labeling which directly uses the lower/upper bound as the label) of \name to train MSCN. 
 
For comparison, we use the OOD queries from Section~\ref{section.exp.card} to evaluate the performances of both variations of \name in MSCN's cardinality estimation accuracy. We also evaluate the ratio of violations of the PK-FK inequality constraint, to which the pseudo-labeling strategy applies.  We report the results on the DSB dataset as we observe that the result on the IMDb dataset shares a similar trend.

\begin{table}[ht]
\centering
\scalebox{0.95}{
\begin{tabular}{|c|c|c|c|c|}
  \hline
  \multirow{2}{*}{Labeling Strategy}  &\multicolumn{3}{c|}{Estimation Accuracy}  & \multirow{2}{*}{\makecell{Ratio of Constraint \\ Violations}}  \\
\cline{2-4}
  & Median & 95th & 99th  &  \\
  \hline
{Labeling by bound}  &  1.3 & 7.7 &  29 &  $5.2\%$   \\
  \hline
   \makecell{Pseudo-Labeling}  & 1.2 &  7.0 & 18 & $4.9\%$    \\ 
  \hline
\end{tabular}}
\caption{Comparison of MSCN w/ and without pseudo-labeling on DSB dataset.}
\label{exp.ablation.label}
\end{table}

Table~\ref{exp.ablation.label} shows the result, from which we can see that pseudo-labeling can help MSCN with better estimation accuracy and slightly fewer constraint violations. But we may notice that their performances in median estimation q-error are very close (1.3 and 1.2), and the improvement mainly focuses on the tail performance --- the 99th-percentile q-errors are 29 and 18 for labeling by bound and Pseudo-Labeling, respectively. 

However, this improvement comes at a cost for the MSCN model, because it has to compute the bitmap results for all $k$ queries generated by the consistency constraint. The process of bitmap look (implemented by Pandas library in Python) can be time-consuming (although it depends on the implementation of bitmap lookup) compared to other constraints that do not require bitmap look such as PK-FK equality constraints and normal SGD steps in training the MSCN model. 
As a result, if not emphasizing the tail performance, we could turn off the pseudo-labeling mechanism in \name for better training efficiency.



\section{Related Work}

The use of constraints in query optimization dates all the way to the original Selinger paper~\cite{selinger1979access}, where index sizes and key-foreign key constraints are among the information used to perform cardinality estimates. Another approach to establishing constraints over costs and cardinalities has been to leverage knowledge from subexpressions computed during execution, and generalizing to others --- for histograms~\cite{aboulnaga1999self, bruno2001stholes, lim2003sash},  query expression statistics~\cite{bruno2002exploiting} and adjustments to correlated predicates~\cite{markl2003leo}.

Over the past decade, approaches based on machine learning techniques have shown great promise.  \textbf{Learned data-driven methods}~\cite{deepdb} do offline computation over samples of existing database instances, to build a model of correlations, selectivities even in the presence of skew, etc.  \textbf{Learned query-driven database systems} can learn or improve an ML model for a variety of database components, by using the execution log of a query workload~\cite{anagnostopoulos2017query,anagnostopoulos2015learning,anagnostopoulos2015learningidcm}.
More recently, there is active work on workload-aware cardinality estimators~\cite{wu2018towards,kipf2019learned,wu2021unified}, cost estimators~\cite{sun2019end, siddiqui2020cost, zhi2021efficient, zhou2020query}, query optimizers~\cite{marcus2022bao,marcus2019neo,kaoudi2020ml}, and workload-aware indexes~\cite{dong2022rw, ding2020tsunami}. Note that we do not aim to provide an exhaustive list in the area.

A relevant cardinality estimation benchmark paper~\cite{wang2020we} also discusses the issue of potential illogical behaviors of learned cardinality models. Our work \name is the first framework that alleviates this issue in query-driven models by introducing domain knowledge into model training.
Two related works aim to improve the query-driven learned cardinality estimation models~\cite{negi2021flow, negi2023robust}, but from different perspectives with different goals. They are both orthogonal to and complementary to this paper.
Specifically, ~\cite{negi2023robust} carefully modifies the MSCN model to achieve better robustness to workload drifts.  Flow-Loss~\cite{negi2021flow} focuses on learning which estimation errors actually matter in plan selection. Interestingly, Flow-Loss can be interpreted as learning domain knowledge from training workloads since the model learns what (sub)queries' errors are important to query optimization, while the philosophy of \name is to explicitly add known domain knowledge (which is valuable in the context of (learned) databases) to a query-driven model for better performance.
As we validate in our experimental analysis, our methods, which incorporate knowledge of database constraints, allow improved cardinality estimation in query-driven systems. It would be interesting to combine \name with both FlossLoss and RobustMSCN~\cite{negi2023robust} as they are compatible --- we can replace the empirical training loss in \name with Flowloss and use it to train the RobustMSCN model. We leave it as future work.
Moreover, we also wish to note that the general framework of \name can be applied to other query-driven learned DB tasks, such as latency prediction~\cite{marcus2019plan} and CPU usage prediction~\cite{saxena2023auto}.

\section{Conclusions and Future Work}

This paper presented \name, a novel methodology for incorporating domain knowledge in the form of \emph{constraints} into learned query optimization.  We demonstrated empirically that existing learned query optimizers perform badly with common classes of constraints used in traditional query optimization; we developed a general framework (including augmenting data and the loss function) to incorporate such constraints into learning; and we showed experimentally that our methods produce substantially better cardinality estimates over the popular IMDb and DSB benchmarks.

We are excited to pursue a number of future directions building upon this work. There is potential for incorporating domain knowledge into full cost (not just cardinality) estimation models; and for incorporating \emph{data-driven estimates} (e.g., from histograms or data-driven models) into our query-driven methods, as a ``soft'' variation on constraints. We also plan to study the question of \emph{how best to sample} from the query workload to generate the most effective training data for constraints.

\clearpage

\bibliographystyle{ACM-Reference-Format}
\bibliography{main}

\end{document}